# Two-Level Concept-Oriented Data Model

Alexandr Savinov

Institute of Mathematics and Computer Science, Academy of Sciences of Moldova
str. Academiei 5, 2028 Chisinau, Moldova
Department of Computer Science III, University of Bonn
Römerstr. 164, 53117 Bonn, Germany

http://conceptoriented.com/savinov/

**Abstract.** In this paper we describe a new approach to data modelling called the concept-oriented model (CoM). This model is based on the formalism of nested ordered sets which uses inclusion relation to produce hierarchical structure of sets and ordering relation to produce multi-dimensional structure among its elements. Nested ordered set is defined as an ordered set where an each element can be itself an ordered set. Ordering relation in CoM is used to define data semantics and operations with data such as projection and de-projection. This data model can be applied to very different problems and the paper describes some its uses such grouping with aggregation and multi-dimensional analysis.

# 1 Introduction

Currently there exist several major data models like hierarchical, network, relational, deductive or multi-dimensional data model each of them having numerous variations or modifications. Each of these major approaches emphasizes one or a few important phenomena or properties of data while other mechanisms are derived or added as a secondary feature. For example, in the hierarchical model the primary property of any data element is that it exists in a hierarchy. Obviously, the presence of hierarchy is one of the fundamental properties which has much wider scope than only data modelling and hence there is no doubt that support of hierarchies should be present in any good data model. However, the problem is that the postulates of the hierarchical data model make it very difficult to develop other very important mechanisms inherent to data modelling. One of such mechanisms is data connectivity which is one of the primary concerns in the network model of data. However, this model of data develops data connectivity features in the prejudice of other important mechanisms one of which is a rich set of operations with data supported in the relational model. The main feature of the relational model in this context is that it deliberately refuses to support any high level mechanism like hierarchies or connectivity but provides a rich set of algebraic operations instead of that. The data modeller in this situation can manually implement any desirable mechanism using a relational query language. There exist also other generic approaches to data modelling like deductive or multi-dimensional. However, they have the same problem of covering only a limited set of mechanisms and hence they are good in one situation and bad in other situations.

In this paper we describe a new approach modelling which is called the concept-oriented data model (CoM) and is based on the previous research results described in [Sav04, Sav05a, Sav05b, Sav06a, Sav06b, Sav08a]. This model is part of the concept-oriented paradigm along with the concept-oriented programming. The goal of this model consists in providing a limited set of basic principles that could be applied to wide range of problems that are encountered in data modelling. In particular, this model can inherently support hierarchies and this makes it similar to the hierarchical data model. It supports connectivity of data elements and in this sense it is similar to the network model. Relational operations are possible in this model and hence it can support rich query languages that can be used to implement a complex logic of data selections and transformations. The concept-oriented model supports inference and constraint propagation – the main features of deductive data models. CoM is inherently multi-dimensional data model which allows the modeller to think in terms of dimensions and in this sense it covers the problems normally described using multi-dimensional models and OLAP techniques.

Although CoM can be viewed as an integrated model combining most important features and mechanisms existing in other models it is not simply a sum of available approaches. Moreover, it does not use the existing approaches for defining its principles at all but rather derives the existing mechanisms from a small set of new general principles. In other words, CoM postulates a small number of general principles independently of the existing models and after that it shows how these principles can be used to model various data modelling patterns and to simulate existing approaches to data modelling.

One of the main concept-oriented postulates is the *principle of duality*. This principle means that in the concept-oriented paradigm any element consists of two parts, called identity and entity. Hence we describe and manipulate pairs rather than single elements. The two sides of any elements are separate but on the other hand compose one whole. Both identity and entity are equally important in data modelling and we cannot ignore any of them. In particular, both identities and entities have structure, behaviour and connections which have to be described by the data modeller using dual facilities provided in CoM.

One consequence of the principle of duality is that data modelling is broken into two orthogonal directions, called *identity modelling* and *entity modelling*. Identity modelling is intended for describing how elements are represented and accessed. The structure of identities is referred to as *physical structure* of data model and it is assumed that all elements exist within a hierarchy. Entity modelling is intended for describing data semantics which depends on how elements (entities) are connected with other elements. The structure of entities is referred to as logical structure and it has a multi-dimensional form where any element has a number of parents and a number of children.

Physical structure of identities is based on *inclusion relation* while logical structure of entities is based on *ordering relation*. In this sense data semantics is based on the formalism of ordered sets. In other words, if elements are ordered in one way then they have one meaning and if this ordering is changed then the meaning is also changed. In order to combine the hierarchical nature of identities and multi-dimensional nature of entities we generalize the formalize of ordered sets and propose to use so called nested ordered sets. The main difference of this new approach from conventional ordered sets is that sets are not flat but rather have a hierarchical structure where any element of the ordered set can also be a set with its own elements having some ordered structure. Using ordering relation and nested ordered sets is an original feature of the concept-oriented model and to the best of our knowledge this approach has not been exploited previously in the research literature. Thus an interesting research challenge is whether it is possible to develop a full-featured data model which is general enough to cover most applications and data modelling patterns.

This paper is structured as follows. In Section 2 we shortly describe what ordered sets are and define one-level concept-oriented data model that is based on this formalism. In Section 3 we introduce the notion of nested ordered set as a basis for the two-level concept-oriented data model. Section 4 is devoted to more thorough investigation of the two-level concept-oriented model including operations with data semantics and representation mechanisms. Section xx describes how this model can be used to solve typical data modelling tasks. Related work is discussed in Section 5 and Section 6 is the conclusion.

# 2 One-Level Data Model

## 2.1 Labelled Ordered Sets

Let us assume that there is a set consisting of a number elements: $O = \{a, b, c, \cdots\}$ (Fig. 1 a). For example, it could be a bibliographic database consisting of three tables, `BIBL = {Author, AuthorTitle, Title}`, where one element is one table (Fig. 1 b). Or it could be one table consisting of records, `Author = {"Immanuel Kant", "Rene Descartes", "Euclid"}`, where one element is one author (Fig. 1 b). Here we do not make any assumptions on the structure and nature of the elements, i.e., all the elements including the set $O$ are things without any internal structure. The only structural fact (except that these elements exist) is that one element, set $O$, includes other elements $a, b, c, \ldots$ (dotted lines in Fig. 1). We will refer to such an inclusion as *physical* or *hard*.

It is important to understand that before elements can be studied they need to exist and the fact of existence is expressed via their *identity*. Identity modelling is one of the most important issues in the concept-oriented approach which is orthogonal to entity modelling. However, in this paper the problem of identity modelling is not discussed (cf. [Sav05c, Sav07a, Sav07b, Sav08b, Sav08c] for more details on how identity can be modelled in programming). Therefore for simplicity we assume that elements are identified using mathematical symbols. Each symbol is unique and has meaning only in the context of its (physical) set. In particular, one and the same symbol can be used to identify elements in different sets. Without this physical inclusion hierarchy we actually have no structure, no meaning, no dependencies and no other non-trivial properties. It is simply a number of non-related elements. For example, table `Title` is not related to table `Author` because the only fact we know is that these two elements exist (in set `BIBL`).

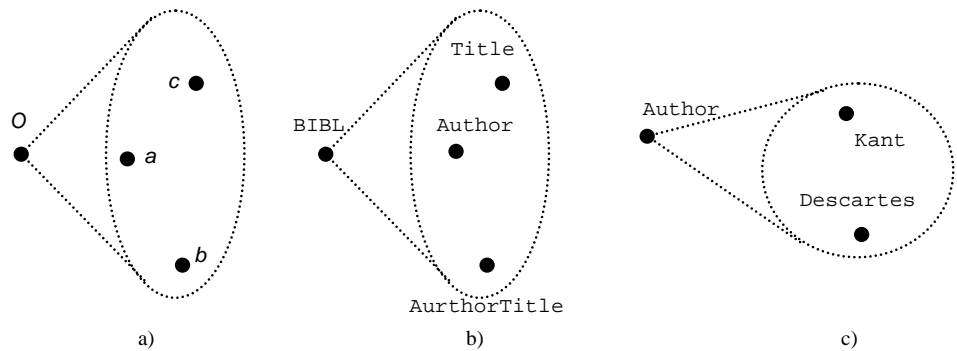

Fig. 1. A physical set without any logical structure.

If we have a set of isolated and unrelated elements the next question is how can we bring a structure into it, which can be then used for modelling by representing data semantics? In other words, we need some formal way for structuring a physical set which is general enough to represent typical situations and patterns accepted in data modelling. The main distinguishing feature of CoM is that for this purpose it uses the mathematical notion of *ordered set*.

Informally, elements of a set are ordered if we can compare them, i.e., for a pair of elements we can say if one of them is less than the other. To compare elements a binary relation is used which called *ordering relation*. Depending on its properties we get different kinds of ordered sets.

*Partial order* is a binary relation '≤' on elements of the set $O = \{a, b, c, \ldots\}$ satisfying the following properties:

[Reflexivity] $\forall a \in O \quad a \leq a$

[Antisymmetry] $\forall a, b \in O \quad (a \leq b) \wedge (b \leq a) \Leftrightarrow a = b$

[Transitivity] $\forall a, b, c \in O \quad (a \leq b) \wedge (b \leq c) \Rightarrow a = c$

Then a *partially ordered set* is a set with partial order established on its elements.

One of the most interesting structures studied in the theory of ordered sets is that of a lattice which is an ordered set satisfying the following additional criterion:

[Lattice] Any two elements $a, b \in O$ have both a least upper bound $\sup(a, b)$ (supremum) and a greatest lower bound $\inf(a, b)$ (infimum).

In mathematics a least upper bound is associated with a generalized product and in different branches has different notation such as $a \cdot b$ (product of two elements), $a \wedge b$ (conjunction of two elements), $a \cap b$ (intersection of two elements). A greatest lower bound is thought of as a generalized sum and is denoted as $a + b$ (sum of two elements), $a \vee b$ (disjunction of two elements), $a \cup b$ (union of two elements).

We also assume that a lattice has two special elements. The greatest element *g* is greater than or equal to any other element of the set: $\forall a \in O \quad a \leq g$. The greatest element is also called *top* and denoted as ⊤. The least element *l* is less than or equal to any other element of the set: $\forall a \in O \quad l \leq a$. The least element is also called bottom and denoted as ⊥.

An element which is greater than this element is referred to as its *super-element*. An element which is less than this element is referred to as its *sub-element*. If $a \leq b$ then *a* is a sub-element for *b* and *b* is a super-element for *a*. In other words, sub-elements are less than this one, and super-elements are greater than this element.

It is convenient to represent an ordered set as a graph where elements are nodes and partial order is represented by directed edges between them. Namely, one edge is an arrow from a smaller element to a greater element. Additionally, we will position a greater element above smaller elements. For example, if $a \leq b$ then an arrow will lead from node *a* to node *b* and *a* is drawn under *b*. Thus all arrows in such a graph are upward directed and lead from sub-elements to super-elements.

An example of an ordered set is shown in Fig. 2 a where element *a* is less than both *b* and *c* and hence *a* is a sub-element for *b* and *c* (while *b* and *c* are its super-elements). Using meaningful names for elements we might represent a model as shown in Fig. 2 b. Here table `AuthorTitle` is a sub-element of both `Author` and `Title`. We see that *a* and `AuthorTitle` in Fig. 2 a-b are bottom elements but top element does not exist. Therefore we simply add it to the set as a special formal element as shown in Fig. 2 c.

In CoM a modified approach to representing ordered sets is used. The difference is that each element of the ordering relation has a unique name which distinguishes it from others. This means that each edge of the ordered set graph is labelled and then it is referred to as a *labelled ordered set*. Such a modification has the following effect. Any two elements can be connected by more than one edge in the ordered set graph having different names. In this case specifying that one element is less than another is not enough. It is necessary to also specify the name(s) of the corresponding ordering relation element(s) as a parameter.

At the same time we do not permit loops in the ordered set, i.e., there is no a sequence of edges in the graph leading from an element to this same element. Consequently, we will use '<' (less than) for denoting ordering relation instead of '≤' (less than or equal to). If two elements are directly connected in the labelled ordered set graph by edge *x* then we write it as follows: $a <_x b$.

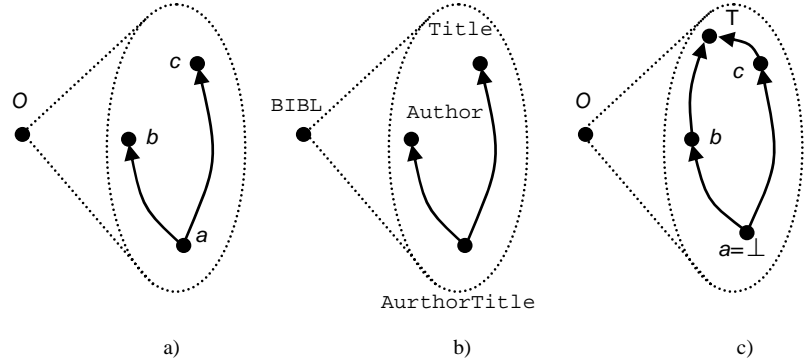

*Fig. 2. An example of ordered set.*

This modification allows us to introduce several definitions important for the whole approach. An edge *x* between two elements *a* and *b*, $a <_x b$, is referred to as a (simple or local) *dimension* of the source element *a*. Given dimension *x* its source is denoted as $\mathrm{Src}(x) = a$. The target of a dimension is referred to as a *domain* and is denoted as $\mathrm{Dom}(x) = b$. (Notice that this notation is accepted in data modelling but differs from mathematics where domain is defined as the source of a mapping.) We will also denote the target (domain) by the dimension itself, i.e., symbol *x* can denote both the dimension itself and (for simplicity) the element *b* where it ends.

The number of dimensions of an element is called its *dimensionality* or intension. i.e., dimensionality of element *a* is the power of the set $\{x_i \mid \mathrm{Src}(x_i) = a\}$. Thus dimensionality is a characteristic of an element which is equal to the number of upward arrows starting from this element and leading to its super-elements. Notice that many dimensions can lead to one super-element and this is precisely why we introduced labels. For example, two dimensions *x* and *y* can start in *a* and end in *b*. In this case dimensionality of element *a* is 2.

The number of dimensions that end in this element is referred to as its (simple or local) *cardinality* or extension, i.e., cardinality of element *b* is the power of the set $\{x_i \mid \mathrm{Dom}(x_i) = b\}$. Thus cardinality is equal to the number of incoming arrows while dimensionality is the number of outgoing arrows. Notice again that a sub-element can be counted many times if there are many dimensions leading to this element.

A *complex dimension* of element *a* is a sequence of local dimensions (separated by dots) where first dimension belongs to *a* and each next dimension starts where the previous dimension ends: $x_1.x_2.\cdots.x_k$, $\mathrm{Src}(x_1) = a$ and $\mathrm{Src}(x_i) = \mathrm{Dom}(x_{i-1})$ where $i = 2,3,\ldots,k$. An *inverse dimension* or sub-dimension is a complex dimension with the opposite direction, i.e., it is a sequence of simple dimensions leading from this element down to some its direct or indirect sub-element: $x_k.\cdots.x_2.x_1$.

The number *k* of local dimensions in a dimension or inverse dimension is referred to as a dimension *rank*. So a simple dimension has rank 1. In an ordered set graph, a complex dimension is an upward path from this element to some its direct or indirect super-element. An inverse dimension is a downward path from this element to some its direct or indirect sub-element.

The number of dimensions from this element to top element is referred to as a *primitive dimensionality* of this element and the number of inverse dimensions to bottom element is referred to as a *primitive cardinality*. A full or *canonical dimensionality* of an element is the number of dimensions to all its direct or indirect super-elements. A full or *canonical cardinality* is the number of inverse dimensions to all its direct or indirect sub-elements.

The same characteristics for the whole labelled lattice are those defined for top and bottom. This means that the lattice primitive/canonical dimensionality is that of bottom element. And the lattice primitive/canonical cardinality is that of top element. In graph terms, the lattice dimensionality is the number of paths leading from bottom to top and cardinality is the number of paths from top to bottom (these numbers are apparently equal).

Direct sub-elements of top element are referred to as primitive elements. We will assume that primitive elements have only one dimension leading to top element.

## 2.2 Interpretations of Ordering Relation

Formally, the only thing that is needed to define a concrete one-level concept-oriented data model consists in specifying some labelled lattice. In other words, we need to define a set of data elements and then order them. This lattice is then considered a full formal specification of the model.

The main practical question is why a labelled lattice can be viewed as a model of data, i.e., how the ordered set can be meaningfully interpreted in a way that can be though of as a concrete data model? Without such an interpretation the labelled lattice (and any other formal specification) will still play a role of mathematical notation, technique or formalism rather than a data modelling approach.

In this section we describe the following main interpretations of the ordering relation which can be directly used in data modelling (Fig. 3):

1. General-specific relation
2. Conjunction-disjunction (logical)
3. Collections-combinations (grouping type)
4. Object-attribute-value (characterization)

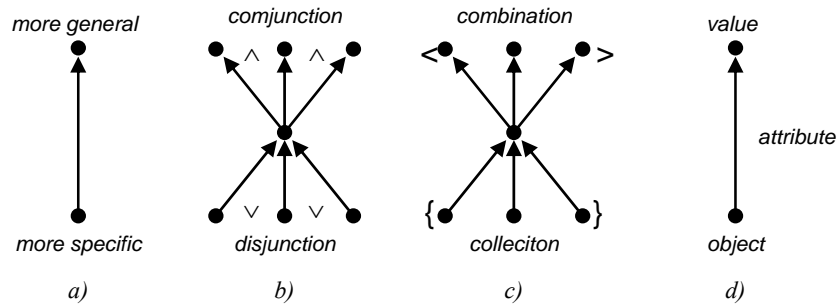

*Fig. 3. Interpretations of ordering relation.*

Frequently data elements of the problem domain can be characterized as more *general* and more *specific* with respect to other elements. For example, in OOP such a description is used to build a class hierarchy where a class is thought of as more specific than its base class the behaviour of which it inherits. In the concept-oriented programming it is inclusion relation that expresses the same semantics of specific-general. The same approach is used in multidimensional databases and OLAP based on describing different levels of details, i.e., more detailed levels have more specific elements.

Ordered sets can be easily used for representing general-specific hierarchy if we assume that an element is more general than its sub-elements and more specific than its super-elements (Fig. 3 a):

$$\forall a,b \in O \text{ if } a < b \text{ then } a \text{ is more specific than } b \text{ and } b \text{ is more general than } a$$

Thus arrows in the ordered set graph lead from a more specific to a more general element. For example, if there exist two elements in the problem domain, `Figure` and `Circle`, and we know that the second one is a special case of the first one then we write it as follows: `Circle<Figure` (circle element is less than figure element). If further we find an element which is even more specific than `Circle` then it is positioned under its parent super-element.

Logical methods are widely used in computer science for problem domain description and analysis. This approach assumes that we can make some propositions about elements which can be true of false. In order to build complex propositions logical connectives are used. For example, if we know that an element must be `Blue` or `Heavy` then we represent this fact using the following logical proposition: `(Blue ∨ Heavy)`. Ordering relation can be interpreted as logical propositions using the assumption that an element is equal to a conjunction of its super-elements and a disjunction of its sub-elements (Fig. 3 b):

$$\forall a,b,c \in O \text{ if } a < b \text{ and } a < c \text{ then } a = b \wedge c$$

$$\forall a,b,c \in O \text{ if } a > b \text{ and } a > c \text{ then } a = b \vee c$$

In the ordered set graph this interpretation can be easily represented if all outgoing arrows are thought of as connected by conjunction while all incoming arrows are connected by disjunction.

In the process of data modelling it is very important to have appropriate means for *grouping*. For example, we might say that some persons belong to a group of employees working on this project. And here again ordering relation can be used to interpret elements as one logical group. However, we distinguish two types of groups called (logical) *collections* and *combinations*. A collection is a group element consisting of its sub-elements. A combination is group element consisting of its super-elements. To distinguish collections and combinations, their elements are written in curly and angle brackets, respectively (Fig. 3 c):

$$\forall a,b,c \in O \text{ if } a < b \text{ and } a < c \text{ then } a = \langle b,c \rangle \text{ - combination}$$

$$\forall a,b,c \in O \text{ if } a > b \text{ and } a > c \text{ then } a = \{b,c\} \text{ - collection}$$

Collections are analogous to conventional sets of elements while combinations are analogous to records, objects or tuples. Thus an element can be thought of as a set consisting of its sub-elements. On the other hand the same element is an object, record or tuple combining in the fields its super-elements.

One of the most wide-spread methods of description is based on using variables which may take different values and characterize the object. In this approach, various phenomena and events are thought of as characterized by some properties or having certain states. For example, an object can be characterized as having blue colour and heavy weight where blue and heavy are values taken by the corresponding characteristics. An ordered set can be interpreted in terms of object characteristics using the assumption that super-elements are *values* characterizing this element while sub-elements are *objects* characterized by this element (Fig. 3 d):

$$\forall a,b \in O \text{ if } a < b \text{ then } b \text{ is a value characterizing object } a$$

According to this interpretation, more specific sub-elements are characterized by their more general super-elements and vice versa super-elements characterize sub-elements. An important property here is that an element can be simultaneously an object (for its super-elements) and a value (for its sub-elements). So we do not have a predefined distinction between objects and values. We cannot say if element `Blue` is a value of some variable or it is an object. For its sub-elements it is a value while for its super-elements it is an object.

Ordering relation allows us to say that some element is characterized by another element. However, values are normally specified in the context of the corresponding variable, which is considered an attribute of one or many objects. In other words, usually we do not say that an object is blue but rather say that the colour of this object is blue where colour is the name of the property. In this case the object-value setting is extended to the object-attribute-value setting. The role of variables or attributes is played by labels of the ordered set, which are named elements of the ordering relation.

$$\forall a,b \in O \text{ if } a <_x b \text{ then } x = b \text{ for object } b \text{ or shortly } a.x = b$$

We say that attribute *x* of object *a* takes value *b*. As we already mentioned, value *b* itself may have its own attributes taking some values and so on till top.

We described four major interpretations of ordering relation but there may be also other interpretations or their minor variations. It is important that formally all these interpretations are equivalent because we use only one source representation as a labelled ordered set. Having different interpretations is useful in model design because this allows us to apply different modelling techniques. For example, we know that a value is always more general than the object it characterizes and a number of values characterizing one object are actually combined using conjunction.

## 2.3 Representation of Labelled Ordered Sets

There are many ways how an ordered set can be represented. For example, it can be written as a set of edges of the ordered set graph where each element is a triple consisting of a sub-element, label name and a super-element. If $a <_x b$ then the triple representing this element of the order is written as $\langle a, x, b \rangle$. This approach is similar to that used in entity-attribute-value model (EAV) and RDF model. Its main property is that by using triples we actually introduce new elements of representation which require their own representation facilities. In other words, the question is how to represent the triples used to represent ordering relation.

This is why we use another approach where elements of the ordered set represent also the order. This approach essentially means that any set is *inherently* an ordered set, i.e., ordering relation is built into any normal set. In other words, if there is a set then it has also an order between its elements as a built-in property. In particular, normal set (without order) is a special case where order is degenerated. Such an approach is a consequence of our general assumption that anything is represented by ordered sets and ordering relation plays primary role in this approach. In terms of the separation between physical and logical structures any set is represented using physical structure while the order is represented using logical structure.

In order to represent an order we need to provide additional characteristics for each element which are defined via other elements from this set. In other words, in addition to an identifier, an element gets some logical characterization which is a specification of the local order. We assume that an element can be defined via its neighbour elements using at least two alternative ways: either by enumerating its super-elements, or by enumerating its sub-elements. Since super-elements are interpreted as more specific or base elements which exist before their sub-elements, it is more natural to use the first alternative where an element is defined via its super-elements.

More specifically, we will assume that each element of an ordered set is a combination of its super-elements:

$$O = \{a, b, \ldots\}, \ a = \langle a_1, a_2, \ldots \rangle, \ b = \langle b_1, b_2, \ldots \rangle, \ \ldots \text{ where } a < a_i, \ b < b_i, \ldots$$

In order to distinguish super-elements within the definition they are be labelled by dimension names:

$$O = \{a, b, \ldots\} \text{ where } a = \langle x_1 : a_1, x_2 : a_2, \ldots \rangle, \ b = \langle y_1 : b_1, y_2 : b_2, \ldots \rangle, \ \ldots$$

This allows us to use one and the same super-element more than once. Such repeating super-elements are distinguished by dimension names which are supposed to be unique for each element. Notice that one element now has two constituents, called *identity* and *entity*. Identity is the element name while entity is a combination of super-element identities. An ordered set is represented as a collection of such identity-entity pairs.

Such a choice of representation is not simply a convenience but reflects a deeper assumption that the fact of writing a number of elements together in one place as a combination defines a new element. In other words, in order to define a new element we need to somehow bring together in one location information on (references to) its constituents (super-elements). Such a combination can be then thought of as an object consisting of a number of fields or a record consisting of a number of column values. It is important however that we cannot separate these values (super-element identities) and store them in different locations because we will loose their unity along with the possibility to interpret them as one whole.

An advantage of such a representation is that elements are defined via elements from this very set, i.e., we do not have elements with some special role. However, a disadvantage is that it is difficult to compare elements because they are defined locally via direct super-elements. In other words, each element has its own local set of dimensions and in this sense they are incompatible. For example, given two arbitrary elements $a$ and $b$ we would like to understand how they are related. Such a possibility to relate and compare elements is very important in data modelling. For example, if it turns out that $a$ is a super-element of $b$ then it can be viewed as its characteristic, attribute value or more general representation. Thus we need a representation or interpretation where elements of the ordered set could be viewed as points in one common space with the structure induced by the ordering relation.

In order to build such a space let us remember that one element is represented as a combination of its super-elements where each super-element can be interpreted as an attribute value (interpretation 4 in section 2.2). This attribute value can be interpreted as a coordinate taken by this element along some dimension or axis. For example, element $a = \langle x_1 : a_1, x_2 : a_2, \ldots \rangle$ has coordinate $a_1$ along dimension $x_1$ and coordinate $a_2$ along dimension $x_2$ and so on. This can be written using dotted notation as follows: $a.x_1 = a_1$, $a.x_2 = a_2$, … The only problem is that these dimensions or axes do not have normal domains from where we could choose possible values (it is a consequence of having one-level model). In this case we can assume that an element has actually the following binary choice: either it has a super-element in its definition (true) or it does not (false). In the former case the super-element is explicitly written in its definition as a coordinate along this dimension (actually the only possible coordinate). In the latter case the super-element simply does not appear in the definition. If $a = \langle x_1 : a_1, x_2 : a_2, \ldots, x_n : a_n \rangle$ then we say that this element is present along *n* dimensions or axes and is absent (not visible) along other possible dimensions. The absent coordinates are simply omitted and we write only the coordinates which are present. If we need to explicitly write the absent super-element then it can be done using a special symbol *null* which is actually a denotation of absence (nothing here). Notice that null is not an element but a marker used to denote the absence of any element along some dimension.

Another important observation is that the definition of an element via its super-elements is actually recursive because the super-elements may have their own definitions. In this case even if this element has one and the same local definition, its relation to other elements may change because its super-elements change their definitions. Thus for relating elements it is important to describe them in common terms which do not change in time.

Let us now consider how elements of an ordered set can be represented as points of one space $\Omega$ with a common set of dimensions. This space consists of all primitive dimensions of the labelled lattice: $\Omega = p_1 \times p_2 \times \ldots \times p_n$, where $p_i$ are primitive dimensions, i.e., dimensions with the domain in some primitive element (a direct sub-element of top element). As we described above a primitive domain here is supposed to have two possible values: either the primitive element itself or nothing (null or false). Space $\Omega$ is then a hypercube where one point is a combination consisting of some primitive elements (a subset of primitive dimensions). An ordered set can be represented as a table with the header corresponding to primitive dimensions and rows corresponding to its elements. We say that the header describes the model *syntax* while the rows represent its *semantics*. For example, let us consider an ordered set shown in Fig. 4. It consists of 8 elements $t, e_1, e_2, \ldots, e_6, b$ and has 6 primitive dimensions (so it is a 6 dimensional model). Hence the table with the primitive semantics will have 6 columns. Each column corresponds to one path from bottom to top as indicated in its header in Fig. 4 by enumerating elements along the primitive dimension path. For example, the first primitive dimension $p_1$ goes through the elements $b, e_1, e_4, t$.

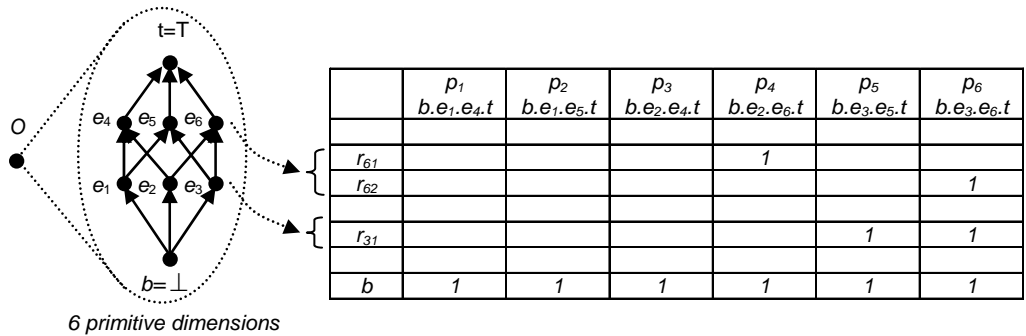

*Fig. 4. Primitive syntax of one-level model.*

Rows are built from elements of the ordered set, i.e., depending on the number of elements and their definition we get one or another set of rows which represent primitive semantics of this syntax. Each element produces many rows. The number of rows for an element is equal to the number of its sub-dimensions leading to bottom element. Equivalently, the number of rows produced by one element is equal to the number of paths leading from bottom to this element. If element $e_i$ has *m* primitive sub-dimensions $f_{ij}$, $j = 1, 2, \ldots, m$ then it produces *m* rows $r_{ij}$, $j = 1, 2, \ldots, m$. For example, element $e_3$ in Fig. 4 has only one sub-dimension $b.e_3$ and hence it is written as one row $r_{31}$ (the first index of the row symbol is the element number from which it is produced). Element $e_6$ has two sub-dimensions $b.e_2.e_6$ and $b.e_3.e_6$, and hence it produces two rows $r_{61}$ and $r_{62}$.

The cells of rows are either 0 or 1 and define the semantics. These values depend on the primitive super-dimensions of the element that produced this row. Formally, if element $e_i$ has $n$ primitive super-dimensions $d_{ik}$, $k = 1,2,\ldots,n$, and $m$ primitive sub-dimensions $f_{ij}$, $j = 1,2,\ldots,m$ then it produces $m$ rows $r_{ij}$, $j = 1,2,\ldots,m$, where one row corresponds to one sub-dimension $f_{ij}$. This row $r_{ij}$ has $n$ columns taking value 1, $p_{ijk} = 1$, which are obtained by concatenating super-dimensions $d_{ik}$ to the sub-dimension $f_{ij}$ (other columns take value 0):

$$r_{ij} = \langle f_{ij}.d_{i1}, f_{ij}.d_{i2}, \ldots, f_{ij}.d_{in} \rangle$$

So each row each is a combination of 1s which correspond to primitive dimensions (columns of the table) where each primitive dimension has prefix $f_{ij}$ and suffix $d_{ij}$. In Fig. 4 element $e_3$ produces one row $r_{31}$ because it has one sub-dimension $b.e_3$. Its semantics is determined by its two primitive super-dimensions $e_3.e_5.t$ and $e_3.e_6.t$. If we concatenate the sub-dimension with these two super-dimensions then we get two primitive dimension paths leading from bottom to top $b.e_3.e_5.t$ and $b.e_3.e_6.t$. These are two last columns in the table which are set to 1 (so it is a two-dimensional element). Element $e_6$ produces two rows $r_{61}$ and $r_{62}$ which both have one super-dimension. If this super-dimension is concatenated with the corresponding sub-dimension the we will get primitive dimensions $b.e_2.e_6.t$ and $b.e_3.e_6.t$, respectively. Hence we set 1 in these columns for these rows.

## 2.4 Operations with Elements

There are two operations that can be applied to individual elements by producing a new element with different dimensions: reduction and extension. The operation of *reduction* removes one dimension from the representation of this element. If the original element has $n$ dimensions, $a = \langle x_1 : a_1, x_2 : a_2, \ldots, x_n : a_n \rangle$, then the reduced element denoted as $a(\rightarrow)x_n$ with the removed last dimension $x_n$ has $n-1$ dimensions:

[Reduction] $a(\rightarrow)x_n = \langle x_1 = a_1, \ldots, x_{n-1} = a_{n-1} \rangle$

Essentially, reduction means that we simply remove one super-element from the definition of this element. Since many super-elements can be used in the definition we use dimension names which are unique.

The operation of *extension* has the opposite form and adds a new dimension to the definition of this element. If the original element has $n$ dimensions, $a = \langle x_1 : a_1, x_2 : a_2, \ldots, x_n : a_n \rangle$, then the extended element denoted as $a(\leftarrow)x_{n+1}$ with the added dimension $x_{n+1}$ has $n+1$ dimensions:

[Extension] $a(\leftarrow)x_{n+1} = \langle x_1 = a_1, \ldots, x_n = a_n, x_{n+1} = a_{n+1} \rangle$

Extension requires also a parameter which specifies the value $a_{n+1}$ of the new dimension. In many cases we can assume that it is *null* which means that the extended element is equivalent to the original one and only formally can be viewed as having the added dimension. However, it is still useful because this allows us to compare elements with different dimensions by formally adding new dimensions.

To compare elements represented as combinations of their super-elements we use the following definition of the induced specific-general relation. Element *a* is more specific (and less general) than element *b* iff they have the same dimensions and each super-element of *b* is present also in *a*. If we assume that marker *null* is more general than any other concrete element, $\forall a, null > a$ then it can be written as follows:

[Specific-general] $a < b \Leftrightarrow a_i < b_i$, $a = \langle x_1 = a_1, \ldots, x_n = a_n \rangle$, $b = \langle x_1 = b_1, \ldots, x_n = b_n \rangle$

Notice that this relation is derived from the definition of elements as combinations of other elements. In other words, if we have a set of records consisting of values then we can induce such a relation on this set. Informally, more general records have more null values than more specific records. Most specific records do not have nulls in their definition and consist of only concrete super-elements. Such records are visible from any dimension.

For example, let us assume that table with primitive semantics shown in Fig. 4 is written as a primary representation rather than produced from an ordered set. The above formulated rule allows us to induce an order between its rows. In particular, row $r_{31}$ is more specific than rows $r_{61}$ and $r_{62}$ because it has 1s where other rows have 0. Row $b$ consists of all 1s and hence it is more specific than any other row. This rule allows us to induce an order on a set of elements with coordinates. However, it is important that the produced ordered set will not be equivalent to the original ordered set just because they have different sets of elements.

# 3 Two-Level Data Model

## 3.1 Nested Ordered Sets

In the previous section we described a one-level model which however has a limited practical use. The main purpose of describing this simplified model consisted in introducing basic notions and definitions concerning ordered sets and demonstrating how this formal setting can be used for data modelling. In particular, we described how ordering relation can be interpreted in conventional terms widely used in data modelling.

One-level data model is very simplified and is not very convenient for describing real world problem domains so it can be hardly used in practice in its direct form. The main problem is that it provides too much freedom for ordering elements while the elements themselves have one and the same status. Indeed, it is very impractical to view a huge number of elements from the problem domain as one big ordered set where an element may have almost any super-elements in its definition without constraints. So the designer has to express the rich variety of real world problems by ordering elements from one big set. In this situation we normally want to introduce some kind of special roles with the corresponding constraints or groups of elements or to apply another kind of structuring facilities.

One of the most widespread practices in data modelling consists in using dedicated groups of elements which can be used only in some context for some purpose. Such groups can be used to restrict the set of elements that can be chosen as super-elements in the definition of another element. In conventional terms a group is interpreted as a domain or class of elements. In such an approach, elements gain an additional structure which is orthogonal to that induced by the ordering relation. This structure assumes that an element has a single parent element which is its permanent group or container and it may have many child elements for which it is a group. This inclusion relation is responsible for describing *physical structure* of the model while the ordering relation is responsible for describing its *logical structure*. In Fig. 5 physical structure of elements spreads horizontally from the root element on the left to leaf elements on the right. Orthogonal logical structure spreads vertically where each element may have a number of super-elements positioned over it and a number of sub-elements positioned under it.

Physical structure is precisely what is studied in the concept-oriented programming (CoP) where it may have any depth and is modelled by a special programming construct, called concept. Physical structure plays a very important role in data modelling. In particular, this distinction between physical structure and logical structure breaks the whole data modelling area into two branches: identity modelling and entity modelling, respectively. In other words, physical structure of the model is where we describe properties and behaviour of identities while logical structure is responsible for that of entities. Yet, even though physical structure plays such an important role, we do not focus on it in this paper and aim at describing only logical structure. We describe only *what* is physical structure but to not touch the issue *how* it can be modelled.

In physical structure, it is assumed that an element is defined as consisting of other elements which in turn may include their own child elements and so on: $e = \{c_1, c_2, \ldots\}$. On the other hand it is still assumed that elements may have a number of super-elements stored as a combination: $e = \langle x_1 = e_1, x_2 = e_2, \ldots \rangle$. Thus an element has actually two definitions: one physical and one logical. One difference between them is that physical structure is permanent and we cannot change the parent of an element which is assigned only once. This assumption is needed because physical structure describes identity of elements which is assumed to be constant, i.e., an element cannot change its identity. In contrast, logical structure may change in time so that an element has one set of super-elements now and another set of super-elements later.

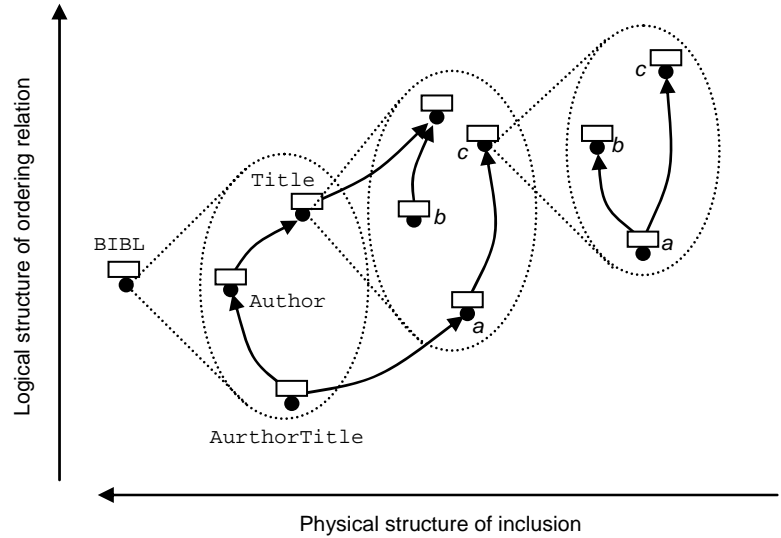

*Fig. 5. Nested ordered set.*

Let us now assume that there is an element which physically belongs to some parent and logically is a combination of some super-elements. The most important new effect of such a definition is that super-elements do not necessarily belong to one set. In other words, if in one-level model super-elements belong to the same set as this element, then in the case of physical structure super-elements may belong to different sets. Such a structure is referred to as a *nested labelled ordered set* and is formally defined below. We still assume that super-elements in logical structure have unique labels and there exist one top element and one bottom element. A nested labelled ordered set is defined as follows:

[Nested labelled ordered set] A nested labelled ordered set is a number of elements where

- each element is defined as consisting of two parts $e = \{c_1, c_2, ...\}\langle x_1 = e_1, x_2 = e_2, ...\rangle$, i.e., an element is:

    (i) [physical structure] a collection of child elements $e = \{c_1, c_2, ...\}$, and

    (ii) [logical structure] a combination of super-elements $e = \langle x_1 = e_1, x_2 = e_2, ...\rangle$

- physical structure is a tree where an element has a single permanent parent and there is one root element $R$,
- logical structure of elements is a labelled ordered set.

If logical structure of the nested ordered set is defined as a labelled lattice (additionally, there are one top and one bottom elements) then this set is referred to as a *nested labelled lattice*. Elements of a labelled ordered set are hierarchically grouped however the groups are normal elements which are also ordered. For example, in Fig. 5 elements `Title`, `Author` and `AuthorTitle` (physically) belong to the root element `BIBL` and hence we write: `BIBL={Title, Author , AuthorTitle}`. These elements can be ordered by storing a combination of their super-elements (shown as upward arrows). What is new in nested sets is that elements themselves can contain child elements which also can be ordered. For example, element `Title` consists of 4 ordered elements and one of these 4 elements consists of 3 ordered elements.

## 3.2 Syntactic Constraints

In a nested labelled ordered set there are no constraints on logical structure except for those specified in the definition. Essentially, elements may have any super-elements provided that there are no cycles. In particular, an element may have super-elements from any level including its parents and children. Here again just as in one-level domain we get two much freedom while the main role of a data model consists in providing mechanisms for imposing structural constraints on its elements. In other words, we have introduced physical structure for restricting the use of elements in logical structure. So if element may still have any super-elements then it does not make much sense to have physical structure. In this section we introduce a mechanism of syntactic constraints which allows us to use physical structure to effectively restrict possible logical structures.

The main idea of the mechanism of syntactic constraints is that possible super-elements depend on the parent of this element. So the position of the element in physical structure determines which logical definition it may have and this is how physical structure constrains logical structure. Formally this principle is formulated as follows:

[Syntactic constraints] If $e = \langle e_1, e_2, \ldots, e_n \rangle \in C = \langle C_1, C_2, \ldots, C_n \rangle$

then $e_i \in C_i$ (directly or indirectly), $i = 1, 2, \ldots, n$.

This principle means that if we have an element $e$ belonging to its physical parent $C = \langle C_1, C_2, \ldots, C_n \rangle$ then the question is how can we define its constituents, i.e., what other elements $e_1, e_2, \ldots, e_n$ can be chosen as its super-elements. Without syntactic constraints we can choose *any* super-elements provided that no cycles appear in the logical structure, i.e., $e_i \in R$ (directly or indirectly super-elements belong to root element). In the presence of syntactic constraints, the parent element $C$ is supposed to be already defined via its super-elements: $C = \langle x_1 : C_1, x_2 : C_2, \ldots, x_n : C_n \rangle$. This definition then is taken into account when defining child elements of $C$. Namely, we assume that any member $e \in C$ may take its super-elements only from within elements $C_1, C_2, \ldots, C_n$. Notice that super-elements of $e$ must belong to super-elements of $C$ directly or indirectly.

Fig. 6 illustrates the principle of syntactic constraints. The root of the model is as usual a single element BIBL which has two child elements Title and Publisher where Publisher is as a super-element for Title. However, by establishing this logical relationship between Title and Publisher we simultaneously impose syntactic constraints on their own child elements. Namely, any child of Title must have super-elements only from within Publisher. For example, the title "Critique of Pure Reason" is defined correctly because its publisher Stiinta belongs to element Publisher which is a super-element of Title which is the parent of this title. The title "Principles of Philosophy" is also defined correctly however its super-element STM is a sub-division of Springer.

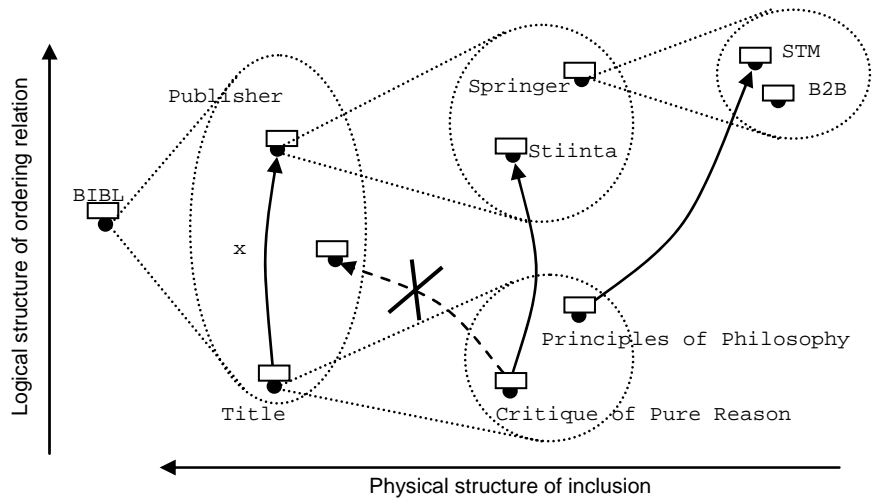

*Fig. 6. Nested ordered set with syntactic constraints.*

By using syntactic constraints parent elements can effectively control how their child elements are defined. Essentially this approach means that any super-element of the parent points to a restricted domain that is allowed to be used by its child elements. In particular, if a parent has no super-elements than its children will not be allowed to have super-elements too.

One of the most interesting interpretations of the formal mechanism of syntactic constraints consists in its use in the object-attribute-value setting. This approach is a generalization of the 4th interpretation of the ordering relation described in section xx. In the one-level model an element can be assigned an arbitrary super-element as one of its properties. However, normally properties of elements are defined for a group of similar elements and have a domain of possible values. For example, fruits are characterized by colour and hence an apple (an element within fruits) can be red or green (elements within colours).

This traditional approach can be easily modelled by the two-level nested ordered set. The two-level model has depth 2 and consists of one root element $R$ (level 0), which physically includes a number of elements called *concepts*, $R = \{x_1 : C_1, x_2 : C_2, \ldots, x_N : C_N\}$, where each concept physically consists of a number elements called (data) *items*, $C_i = \{e_{i1}, e_{i2}, \ldots, e_{im_i}\}$. (For comparison, in one-level model the root consists of data items only: $R = \{e_1, e_2, \ldots, e_N\}$.) Logical structure remains the same, i.e., any element has a number of super-elements: a concept is a combination of super-concepts, $C = \langle C_1, C_2, \ldots, C_n \rangle \in R$ and an item is a combination of super-items, $e = \langle e_1, e_2, \ldots, e_n \rangle \in C$. The difference is that we assume that syntactic constraints are always imposed. This means that concepts may have any structure according to one-level model while data items may take its super-items from only super-concepts of their parent concept: $e_i \in C_i$, $i = 1,2,\ldots,n$.

In the two-level model, concept dimensions can be thought of as properties of internal elements which may take different values as elements of some other concepts. For example, if concept $C$ has super-concept $C_i$ labelled by dimension $x_i$, $C = \langle \ldots, x_i : C_i, \ldots \rangle \in R$, then we say that element $e \in C$ has property or attribute $x_i$ taking value $e_i \in C_i$. In this case the same can be written as follows: $e.x_i = e_i$. Thus all possible properties of the two-level model are described in the logical structure of concepts while items inherit and use this structure.

It is important to understand that attribute-value setting is only an interpretation and all model properties are defined by the ordering relation over nested structure of sets. However, this interpretation allows us to connect real world descriptions in terms of attributes and values with the formal representation using ordered sets. One interesting property of such an approach is that an item can be both a characterized object and some object property value. More specifically, an item is a characterized object for its super-items and it is a value of some property for its sub-items.

All the interpretations described in section xx are also valid for the two-level model. In particular, super-concepts and super-items are considered more general (less specific) than sub-concepts and sub-items. A concept/item is a logical conjunction of its super-concepts/super-items. And a concept/item is a collection of its sub-concepts/sub-items. An interesting consequence of these interpretations is that a value item is more general than the object item it characterizes. And vice versa, an object item is more specific than value items that characterize it.

### 3.3 Multidimensional Hierarchical Space

A flat multi-dimensional space consists of a number of axes or dimensions $x_1, x_2, \ldots, x_n$ where each axis takes its values from a set of values or coordinates called also domains: $x_i \in X_i$, $i = 1,2,\ldots,n$. The Cartesian product of these sets of values is the multi-dimensional space, $\Omega = X_1 \times X_2 \times \ldots \times X_n$, which consists of a number points, $\omega \in \Omega$, where each point is a combination of values, $\omega = \langle x_1, x_2, \ldots, x_n \rangle$. We will say that choosing axes with their values defines space syntax or syntactic structure. If we choose some subset of points in the space then we define semantics or semantic structure. For example, semantics could be defined as a line or hype-place or a more complex surface using some type of equation.

One of the interesting features of the two-level model is that it can be interpreted as a multi-dimensional *hierarchical* space, i.e., a multi-dimensional space which allows for its dimensions to be defined hierarchically. Such a geometrical interpretation is very important for understanding the essence of the concept-oriented model. Shortly, concept structure of the model defines its coordinate axes while items are points in this space. We also refer to the concept structure as model syntax while items define model semantics. Creating a two-level model can be viewed as defining the space syntactic structure which provides constraints for the items as described in the previous section. After that we can define model semantics by adding items which take their coordinates along this space axes. Such an analogy is rather fruitful because thinking in terms of space and points is a very natural and intuitive method.

Concept $Z = \langle x_1 : X_1, x_2 : X_2, \ldots, x_n : X_n \rangle$ can be viewed as a (flat) multi-dimensional space having $n$ dimensions $x_1, x_2, \ldots, x_n$ with domains in super-concepts $X_1, X_2, \ldots, X_n$. This means that each point $z \in Z$ in this space is characterized by $n$ coordinates along the axes where each coordinate is some super-item: $z = \langle x_1 = z_1, x_2 = z_2, \ldots, x_n = z_n \rangle$, $z_i \in X_i$, $i = 1, 2, \ldots, n$. Points from this space may take values only along their domains because of syntactic constraints. However, in contrast to the conventional (flat) multi-dimensional space, points may take no coordinates at all which can be marked by *null*. Super-concepts $X_1, X_2, \ldots, X_n$ are called domains or ranges for dimensions $x_1, x_2, \ldots, x_n$ which is denoted as $\mathrm{Dom}(x_i) = X_i$.

For example (Fig. 7), if we have two coordinate axes $x$ and $y$ with the domains in $X$ and $Y$ then they can be used to build a two-dimensional space $Z$ using two-level model. Domains $X$ and $Y$ are super-concepts of $Z$ labelled by dimensions $x$ and $y$: $Z = \langle x : X, y : Y \rangle$. Points in space $Z$ are items consisting of two super-items taken from $X$ and $Y$. In Fig. 7 each axis has two coordinates (black circles) that can be taken by points from $Z$. Syntactically, space $Z$ can consist of 4 points defined as all combinations of items from $X$ and $Y$ each having two elements. However, semantically, in Fig. 7 we chose only 3 points (shown as black circles). Fig. 7 a is a traditional representation where two dimensions are shown as axes with points. Fig. 7 b-c illustrate how the same space is represented using two-level concept-oriented model. One of the most important features of our approach is that the space is represented using ordering relation, i.e., we show that a nested ordered set can be interpreted as a traditional multi-dimensional space.

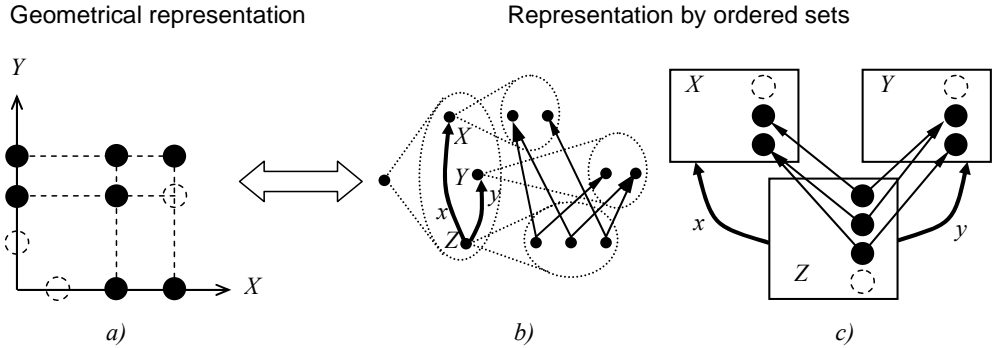

*Fig. 7. Representation of multi-dimensional space by a two-level ordered set.*

An important property of using ordering relation for representing multi-dimensional space is that the role of axis with coordinates and space with points is not strictly assigned. One and the same concept is interpreted as an axis for its sub-concepts and as a multi-dimensional space with respect to its super-concepts. In the syntactic structure all concepts have the same rights and we do not define their concrete role except for having a relative position to other concepts. One consequence of using such space description is that an existing multi-dimensional space can be used as a dimension for other spaces represented by its sub-concepts.

We can bring hierarchy into space structure by extending it downwards. For example, concept $Z$ defined in Fig. 7 as a two-dimensional space with two axes $X$ and $Y$ can be used as a dimension for its sub-concept $W$ (Fig. 8 a). Thus space $W$ has two direct dimensions $Z$ and $U$ and several indirect dimensions. What is new here is that dimension $Z$ has its own internal structure, namely, it is a two-dimensional space.

On the other hand, if we have already an axis then its coordinates are not necessarily primitive items and may have their own internal structure. Such a complex axis is defined as a multi-dimensional space with its own axes. For example, in Fig. 8 b axis $Y$ which was a primitive dimension in Fig. 7, is defined as a two-dimensional space with axes $U$ and $V$. Thus we can bring hierarchy into a space structure by extending it upwards.

By adding new sub-concepts or super-concepts we can increase the depth of the space and add more levels into its structure. So each point still has some coordinates but these coordinates are now points with their own coordinates.

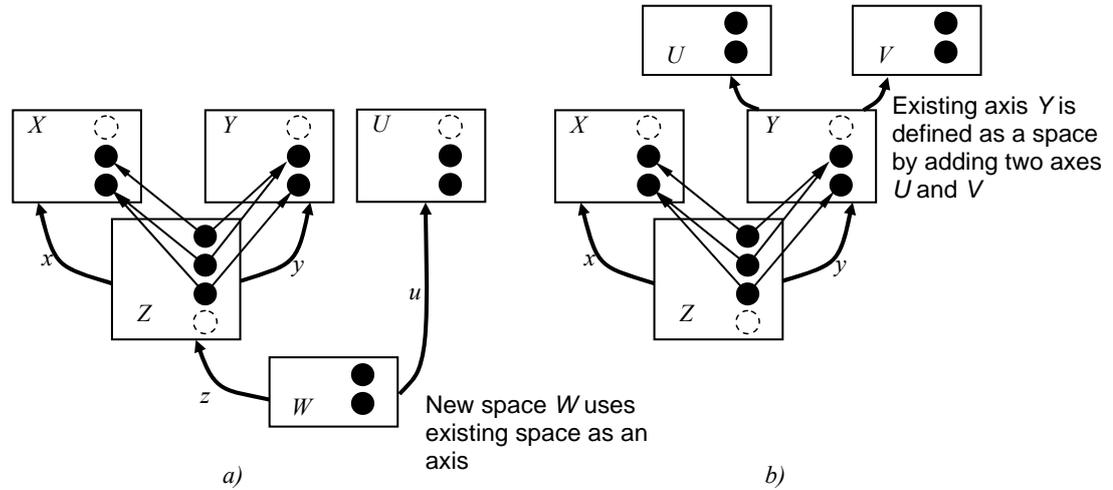

*Fig. 8. Hierarchical multi-dimensional space.*

Such a geometrical interpretation is very natural and, moreover, it is analogous to one wide-spread modelling pattern called snow-flake schema. Indeed, if we rearrange concepts shown in Fig. 8 and draw them as a graph as shown in Fig. 9 then we can easily see that concept *W* (Fig. 9 a) and concept *Z* (Fig. 9 b) play a role of master table while their sub-concepts are detail tables. The main difference of the conventional snow-flake schema pattern is that it does not use ordering for describing the structure of the model and such a schema is a graph which is useful for visualization purposes. Representation by nested ordered sets used in CoM allows us to develop many important mechanisms which are described later on in the paper.

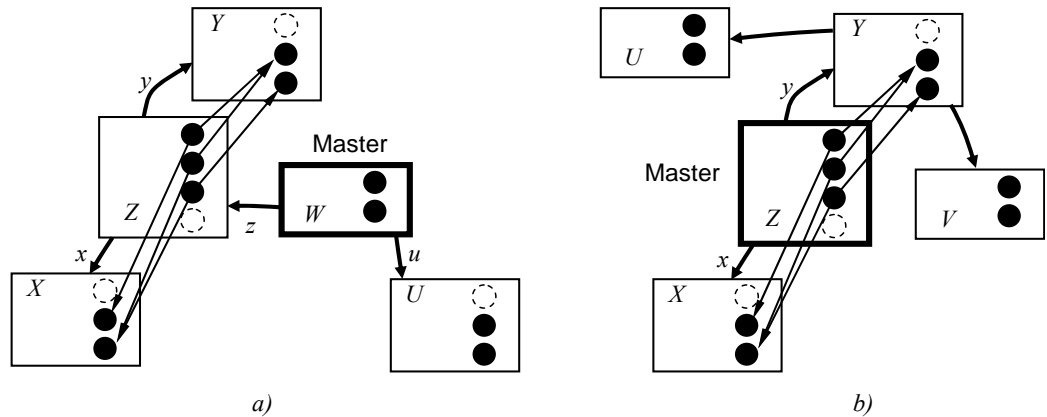

*Fig. 9. Representation by snow-flake schema.*

## 4 Operations with Model Semantics

### 4.1 Representing Model Semantics

Canonical semantics is a representation which allows us to work with different models, compare them and apply operations to them. Normally for such a representation we choose some common terms so that all models are represented as points in one and the same space. The problem is that models are defined using local relationships because it is the way how they are usually built in the problem domain. For example, two models could be defined as two sets of entities with certain relationships within each set. In this case it is not clear how do they relate to each other. In particular, we do not know if one of them is more general than the other and we cannot determine if they are equal. In such a situation we say that this approach does not possesses canonical semantics because we are not able to represent these models using some common terms.

Another important role of formal canonical representation of data semantics is that it allows us to manipulate the whole model as one element. In particular, we can apply operations to the while model rather than to its constituents like data items or concepts. And then different states of one model could be semantically compared. For example, we could say that the next state is more general than the previous one. For example, if we delete a record from one table and add a record to another table then how the meaning of the whole database will change? If we had formal canonical representation then we could easily answer this question.

It is rather important problem which exists actually in many other branches of computer science and other domains. For example, in mathematics this approach allows us to work with functions as elements of some universal space where one function is one point rather than a number of constituents. Depending on the properties of this space we can compare these functions and apply different operations to them. For example, two functions defined indirectly could be actually very similar or they could have some very special relationships between them or they could possess some interesting properties which are derived from their position in the space of functions. In computer programming having formal semantics for a program makes it possible to solve many difficult problems. Indeed, if we have two programs then how can we decide if they are equivalent or not? The only way consists in describing them in some common terms in such a way that all programs are represented as points of one space. Having such a representation we can formally reason about programs by capturing their important aspects and properties. And we can formally prove that a program really solves some problem, i.e., reaches some state or gets some result.

One simple approach to introducing formal semantics into two-level model consists in using all data items from all concepts as a common representation. In this case it is assumed that the structure of the model is always the same, i.e., we can only compare models having the same syntax. However, such a representation is actually rather restrictive. First of all, models could be restructured while their semantics remains the same. A typical example is where one and the same data is represented either in one table or distributed among many tables but these representations are equivalent. Another serious problem is that we cannot actually compare models modified using local operations. For example, let us suppose that we have two models with one and the same syntactic structure. Then we add an item into one model and remove an item from the second model. Will these models change significantly or may be they will be equivalent? We cannot answer this question because each item in the representation uses its local space and hence all operations are defined locally. The main problem is that the multi-dimensional hierarchical structure prevents us from representing models in common terms. In other words, we represent data items using terms (super-items) which themselves have their own representation in other terms.

In order to solve this problem with local terms which do not allow us to work with data items globally we can represent all data items (i.e., all the available model semantics) in only primitive terms which do not have their own definition. These primitive terms will be common to all data items and hence all elements of the model will be represented in one and the same space. As primitive common terms we choose items from primitive concepts which are called *primitive items*. If we can express all data in the model in terms of primitive items then this would allow us not only to compare different models with the same syntax but also to work with models having different syntactic structure. Indeed, the only requirement is that the set of primitive items is the same while the hierarchical multi-dimensional structure based on them can be different. Such a representation is referred to as *primitive semantics* of two-level model and it is analogous to the primitive semantics of one-level model described in section xx.

Primitive semantics of two-level model is defined analogously to one-level model. The idea is that the hierarchical multi-dimensional space is converted into a flat multi-dimensional space consisting of only primitive dimensions. The main difference is that in one-level model primitive dimensions are binary so that a variable either takes its value (primitive item) or does not take it (no value or null). In two-level model primitive dimensions have normal domains which are primitive concepts where values are primitive items. Then variables take values from the primitive domains (absence of value is also possible).

In primitive representation, instead of the hierarchical syntactic structure of the original model we use common space $\Omega = P_1 \times P_2 \times \ldots \times P_N$ where $P_1, P_2, \ldots, P_N$ are domains of primitive dimensions $p_1, p_2, \ldots, p_N$: $p_i \in P_i$, $\text{Dom}(p_i) = P_i$, $i = 1, 2, \ldots, N$. In other words, $p_1, p_2, \ldots, p_N$ are primitive dimensions of the model (paths from bottom to primitive concepts) while $P_1, P_2, \ldots, P_N$ are their domains (primitive concepts). Since several primitive dimensions of the model may have the same primitive concept as a domain, the list $P_1, P_2, \ldots, P_N$ may include the same primitive concepts. For example, a model might have only one primitive concept $P$ which is used by $N$ primitive dimensions of bottom concept and then the primitive syntax would consist of $N$ concepts $P$: $P_1, P_2, \ldots, P_N$, where $P_i = P$, $i = 1, 2, \ldots, N$.

The representation consisting of all primitive dimensions is inherently flat and therefore it can be naturally represented by a table where columns correspond to primitive dimensions. The table then has $N$ columns $p_1, p_2, \ldots, p_N$ and the row cells take values from concepts $P_1, P_2, \ldots, P_N$. Thus the table with the primitive representation has as many columns as bottom concept of the model has primitive dimensions. The difference from the corresponding representation in one-level model is that now cells take values from primitive concepts as primitive items rather than binary values.

Let us consider a model shown in Fig. 10 consisting of 6 concepts $U$, $V$, $W$, $X$, $Y$, $Z$ where the first three concepts are primitive, i.e., are direct sub-concepts of top concept (not shown). In order to determine the structure of the table with primitive semantics we can ignore data items and consider only concept structure. The primitive syntax of this model has 5 dimensions leading from bottom concept $Z$ to primitive concepts $U$, $V$, $W$. Hence the table will have 5 columns and each row will have 5 cells taking values from the corresponding primitive concepts. Notice that columns 2-3 take values from one concept $V$ and columns 4-5 take values from one concept $W$.

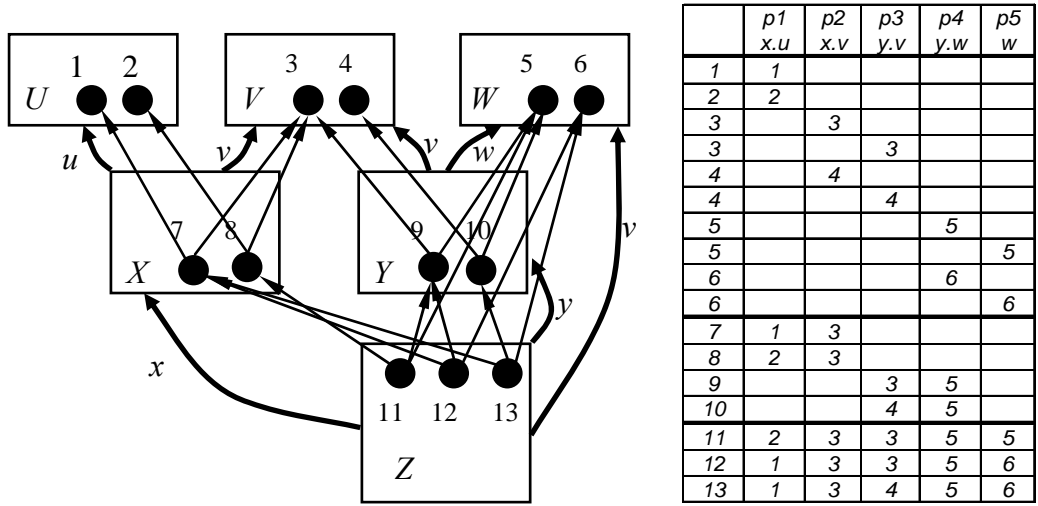

Fig. 10. Primitive semantics of a two-level nested ordered set.

Further we need to convert model semantics into the primitive representation where all items are represented by rows in the space of primitive dimensions. In other words, each item from the original model represented by its local dimensions (super-items) has to be somehow represented as a point in $N$-dimensional space of the model primitive dimensions. Here we use the same approach as for one-level model. Namely, for each original item we insert as many rows as it has sub-dimensions leading to bottom concept. Formally, if item $e$ has $n$ primitive super-dimensions $d_k$, $k = 1, 2, \ldots, n$, and $m$ primitive sub-dimensions $f_j$, $j = 1, 2, \ldots, m$, then it produces $m$ items $r_j$ in the primitive semantics (rows in the tables with primitive columns):

$$r_j = \langle f_j.d_1, f_j.d_2, \ldots, f_j.d_n \rangle$$

This operation can be viewed as extending item $e$ down to bottom concept along $m$ sub-dimensions:

$$r_j = e(\leftarrow) f_j$$

Item extension retains its original super-dimensions $d_1, d_2, \ldots, d_n$ and adds missing primitive dimensions of the target primitive model. In the table the values of new extended dimensions are absent and can be marked by null.

Model shown in Fig. 10 has 5 primitive dimensions (syntax) and 13 items (semantics) including 6 primitive items from concepts $U$, $V$ and $W$. Each item has to be written in the table with primitive semantics as many times as it has sub-dimensions. For example, item 1 is written only once in the first row because it has only one sub-dimension $Z.x.u$. However, item 3 is written two times because it has two sub-dimensions $Z.x.v$ and $Z.y.v$. Items 4, 5 and 6 also produce two rows each because there are two different paths from bottom concept $Z$ to each of them. All the rest of items produce one row. For example, item 7 has two super-dimensions in primitive concepts $U$ and $V$ and hence it is distinguished from other items only by these two values which are written in columns 1 and 2. Its value along first dimension $Z.x.u$ is equal 1 and the value along second dimension $Z.x.v$ is equal 3. So we write these two values (primitive items) in first two columns of the table.

Primitive semantics allows us to think of items as points in a flat multi-dimensional space consisting of primitive dimensions. In particular, using such a representation it is easy to define *specific-general* relation. Item $a$ is more specific (less general) than item $b$ if for any primitive dimension in $b$ it has the same non-null components:

[Specific-general]   $a < b$   $\Leftrightarrow$   $\forall p_i$, $a.p_i = b.p_i$ or $b.p_i = null$

Here $p_i$ is a dimension in the primitive representation of these items (compare it with the definition of specific-general relation for elements in one-level model). Thus the more coordinates a point has, the more specific it is. The most general is an item without coordinates at all, i.e., with all nulls as its super-items. This item is precisely what we mean by top item which is formally supposed to be the only available super-item for any primitive item. The most specific item does not contain nulls at all.

The specific-general relation is useful for defining *coverage* relation. We say that an item covers all its more specific items. In the primitive representation coverage is a number of rows which have the same values as in non-null components of the selected item. For example, item 7 in Fig. 10 has two non-null components in first two columns: $Z.y.u=1$ and $Z.y.v=3$. It therefore covers two rows 12 and 13 which have the same values in these columns.

It is important that primitive semantics can be used to determine relationships among elements (specific-general, coverage etc.) only if all items are semantically different. If there are two items having one and the same super-items (the same semantic definition) then they are indistinguishable in the primitive representation. As a consequence these two items are represented by equal rows. What is even worse, their sub-items are also indistinguishable on these components. In this case coverage is larger than it has to be. This property is quite natural because points with the same coordinates are also indistinguishable. When using primitive semantics we will always assume that all items are semantically different. In practice it is not important because all operations are defined in terms of unique identifiers (item references) and hence we can guarantee distinguishablity using unique identifiers.

### 4.2 Projection and De-Projection

The multi-dimensional hierarchical structure provides natural means for navigating over the model. If we have one or more data items from some concept then we can use dimensions and sub-dimensions to retrieve related items. Thus the structure of dimensions provides not only syntactic constraints by restricting possible super-items for an item but it also can be used to move up and down to related super-items or sub-items, respectively.

Informally, if we have an item then getting its super-item along some dimension is thought of as a projection while getting its sub-items along some sub-dimension is thought of as a de-projection. Formally, if $e$ is an item from concept $C$, $e \in C$, and $d$ is some dimension of concept $C$ with domain in $D$, $\text{Dom}(d) = D$, then operation $e \to d$ is referred to as a *projection* of $e$ to $D$ along $d$ and returns super-item $s \in D$:

$e \to d = s$, where $e = \langle \ldots, d : s, \ldots \rangle$ (or $e.d = s$)

So projection of a single item means moving up along the specified dimension and getting a super-item. It is possible that several items have one and the same super-item, i.e., they are projected to one point. It will be assumed that if projection is applied to a set of items then the result does not include repeating super-items. If $E$ is a set of items from $C$, $E \subseteq C$, then their projection along $d$ returns a set of their super-items each taken only once:

$$E \to d = \bigcup_{e \in E} \{s \mid s = e \to d\}$$

If it is necessary to return all super-items for each source element (even if they are the same) then we will use dot instead of arrow to denote this operation. So $E.d$ is a set of super-items from $D$ with as many elements as in $E$: $|E| = |E.d|$ (for comparison, $|E| \geq |E \to d|$). For example, in Fig. 11, the result of projecting three items of set $I$ to super-concept $D$ is two items: $I \to d = \{2,3\}$. If three items of concept $F$ are projected to concept $D$ then we will get one item: $F \to f \to d = \{3\}$. Notice that we can get repeating items using dot operation: $F.f.d = \{3,3,3\}$. We can also mix dots and arrows, for example: $F \to f.d = \{3,3\}$.

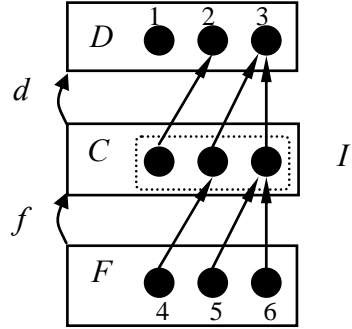

*Fig. 11. Projection and de-projection.*

If $e$ is an item from concept $C$, $e \in C$, and $f$ is a dimension of its sub-concept $F = \langle ..., f : C, ... \rangle$ with the domain in $C$, $\mathrm{Dom}(f) = C$, then operation $e \leftarrow f \leftarrow F$ is referred to as a *de-projection* of $e$ to $F$ along $f$ and returns a set of sub-items with projection in this item:

$$e \leftarrow f \leftarrow F = \{s \mid s \to f = e\}$$

For example (Fig. 11), if items from set $I$ are de-projected to sub-concept $F$ along $f$ then we will get three items as a result: $I \leftarrow f \leftarrow F = \{4,5,6\}$. If items 2 and 3 from concept $D$ are de-projected to $F$ then we will also get three items: $\{2,3\} \leftarrow d \leftarrow f \leftarrow F = \{4,5,6\}$.

Notice that for projection we need to specify only the source element(s) and one dimension because the domain is unambiguously determined by the dimension. For de-projection we need to additionally specify the target sub-concept as an additional parameter of the operation. It is because a concept may have many sub-dimensions with one name and these de-projection directions need to be somehow distinguished.

Another important point is that dimensions used to project or de-project elements may have any rank. This means that we can find a projection not only to a direct super-concept but also to any indirect super-concept along a longer upward path. In the graph of the ordered set finding projection of an item means moving up along the path with the specified name and retrieving the target super-item. Finding de-projection means moving down along the specified path and retrieving all the target sub-items.

Full de-projection (along *all* sub-dimensions of any rank) is a set of all more specific items in the model $R$: $\{s \in R \mid \exists e \in I, s < e\}$. In the table with primitive semantics, the full de-projection is a set of rows covered by these items. (As we already noticed, this approach for finding de-projection works only if all items are semantically different, i.e., the model does not have two items with the same super-items.)

Projection and de-projection operations have a clear geometrical interpretation. If this concept is a multi-dimensional space then its super-concepts are axes. Any item from this concept has coordinates along these axes and hence these coordinates are its projections on these axes. In other words, if we look at an item from some of its super-concepts then we will see only its super-item (if it has null along this axis then we will see nothing, i.e., the item is invisible along this axis). In data modelling this operation allows us to narrow down the view by projecting any data item along any its dimension. An interesting observation here is that projection of one item is equivalent to getting its attribute values, i.e., attribute values of an object are its projections along the corresponding properties. Thus properties are views of this object from different dimensions.

Notice also that projection is analogous to the operation of reduction described in section xx. Both operations result in an element with fewer dimensions. The difference is that reduction is applied to this element and produces a new element independently of the existence of other elements in the model (particularly, independently of the existence of super-elements). In contrast, projection returns an existing element from this model which is somehow related to this element. More specifically, it returns a super-element given a dimension. The same analogy exists between extension and de-projection operations. The former transforms the argument and returns some result in any case while the latter returns a set of already existing sub-elements from this model.

One of the most interesting extensions of the described projection and de-projection operations is that they can be applied consecutively by producing a chain of upward and downward segments in the concept graph. Such a sequence of projection and de-projection operations where each next operation is applied to the result of the previous operation is referred to as a *logical access path*. A concept where the access path changes its direction on the opposite one is called a *turning point*. For example, in Fig. 11 item 1 from concept $F$ can be projected along dimensions $f$ and $d$ to concept $D$ and then the result is de-projected back to concept $F$ in the opposite direction. The result of this access path consists of three items $\{4, 5, 6\}$ from concept $F$:

$$\{4\} \to f \to d \to \overline{D} \leftarrow d \leftarrow f \leftarrow F = \{4,5,6\}$$

Here concept $D$ (marked by bar) is a turning point which can be distinguished by the presence of two incoming arrows (from left and right). If the obtained set is projected again to concept $D$ then we will get item 3:

$$\{4\} \to f \to d \to \overline{D} \leftarrow d \leftarrow f \leftarrow \overline{F} \to f \to d = \{3\}$$

Here we have two turning points $D$ (upper) and $F$ (lower) because the path changes its direction two times.

Assume that a concept has many dimensions leading to some its super-concept. In this case we could select super-items which are referenced by selected items from this concept along all these dimensions. Such a projection is referred to as a *multi-dimensional projection*. Obviously, the result of multi-dimensional projection is equal to intersection of projections along individual dimensions:

$$E \to \langle d_1, d_2, \ldots, d_n \rangle = \bigcap_{i=1,2,\ldots,n} E \to d_i$$

Here $E$ is a subset of items from $C$ and $\langle d_1, d_2, \ldots, d_n \rangle$ is a set of dimensions along which we want to project all having one super-concept (one domain), $\forall i, j : \text{Dom}(d_i) = \text{Dom}(d_j) = D$.

Multi-dimensional de-projection is defined analogously. In this case we want to find all sub-items which reference these items along all selected sub-dimensions.

$$E \leftarrow \langle f_1, f_2, \ldots, f_m \rangle \leftarrow F = \bigcap_{i=1,2,\ldots,m} E \leftarrow f_i \leftarrow F$$

## 4.3 Constraints and their Propagation

Constraints are widely used in data modelling because in most cases we need to select only part of the whole model which satisfies certain conditions. For example, we might want to select only objects with certain properties, say, persons with some birth date. However, in the concept-oriented model we formally have only a nested set of ordered elements and hence we need some method for imposing constraints on it and selecting a subset of this ordered set of elements. Here we have two major problems. One is that elements are organized into a multi-dimensional hierarchy and hence this set cannot be considered as a flat multi-dimensional space. The second problem is that constraints are normally imposed locally by specifying properties of items but we need a method for their propagation through the whole model. In other words, it is necessary to understand how local constraints influence elements in other parts of the model.

Let us assume that set *O* also called the universe of discourse does not possess any structure and consists of all *syntactically correct* elements. Essentially, each element $o \in O$ is a potentially possible data item that we can construct and use in such a trivial model. Further, we select subset *M* of this set *O*, $M \subseteq O$, which consists of all *semantically correct* elements, i.e., elements which are considered meaningful in the problem domain. Subset *M* reflects the current state of the problem domain and we can change it by adding or removing elements. There exist numerous ways for representing subset *M* of meaningful elements, e.g., analytically using some equation, logically using predicate calculus or explicitly by enumerating all its elements. But any approach has to provide means for determining semantic value for any element $o \in O$. This semantic assignment is actually a *characteristic function* or distribution from the set *O* of all syntactically correct elements to the set *L* of semantic values: $\varphi : O \rightarrow L$. If we represent this characteristic function then we essentially represent all the data (or knowledge) in our system.

Set *L* may have different composition and structure. For example, if it consists of all points from the interval [0,1] then we can build a system with fuzzy semantics or probabilistic semantics (depending on the set of operations). However, one of the simplest cases is where it consists of only two elements 0 and 1, $L = \{0,1\}$, which are interpreted as false and true, respectively. We say that in this case each element from the universe of discourse *O* can be either meaningful or not. In databases set *M* is normally stored explicitly by creating the corresponding objects identified by their references. In this case semantic values from *L* get existential semantics, i.e., if an object exists in the database then we say that it belongs to set *M* and is meaningful. Otherwise, if it does not exist in the database then we say that it is assigned value 0. Yet we can explicitly assign existence as a marker. If an existing object is marked by 1 then it is assumed to really exist. If it is marked by 0 (or null) then we assume that it does not exist.

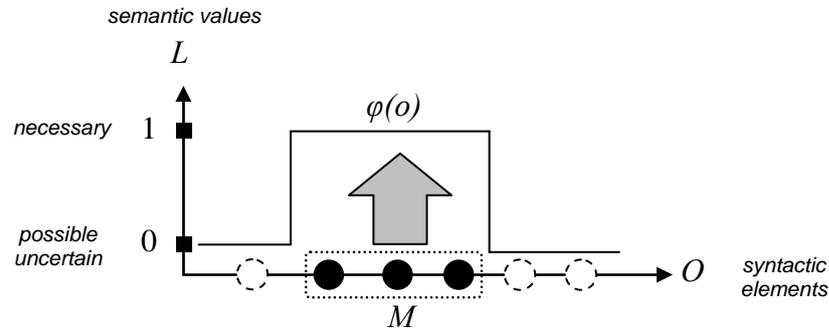

*Fig. 12. Necessity semantics.*

Two semantic values 0 and 1 may have different interpretations. One approach assumes that a data item exists in the database (marked by 1) only if we are absolutely sure that it exists in the real world (Fig. 12). For example, if data semantics is represented by subset *M* then we can add a new data element $e \in O$ to it by producing a new state of the database: $M \cup \{e\} = M'$. In this way we say that element *e* is known to exist in the problem domain or is known to be meaningful. *M'* is a new state of the database which includes element *e*. What about other elements which are assigned 0 in our semantics? They are interpreted as uncertainty, i.e., if an element is assigned 0 (is not in the database) then we assume that we do not have any information on it. Thus we know only what we have explicitly while everything else is supposed to be unknown. Semantic value 1 is also called *necessity* (i.e., existing elements necessary exist) while value 0 is called *uncertainty*. Notice that this interpretation assumes that elements marked by 0 are still possible.

The semantics of necessity which used to represent data has one important property: we are not able to represent negative information, i.e., we cannot represent what is known not to exit. Indeed, value 1 is used to mark elements which are known to exist while value 0 is understood as uncertainty. However, frequently our knowledge about the problem domain has a negative character, i.e., we know what is not possible rather than what is necessary. An alternative approach consists in interpreting 1 as uncertainty or possibility while 0 is interpreted as *impossibility*. In other words, elements which are known not to exist (also called disabled or prohibited elements) are assigned semantic value 0 (Fig. 13). This semantics of possibility is used in describing constraints. Thus constraints have always negative character by excluding some data or specifying what is *not* interesting in the current context. Such prohibited intervals are also called holes in data. Constraints can be also expressed via a function or distribution from the set of all syntactically correct elements to the set of semantic values: $\psi : O \rightarrow L$. However, elements of set $L$ have the semantics of possibility rather than semantics of necessity. As a result, if we combine two constraints then the number of excluded elements will increase.

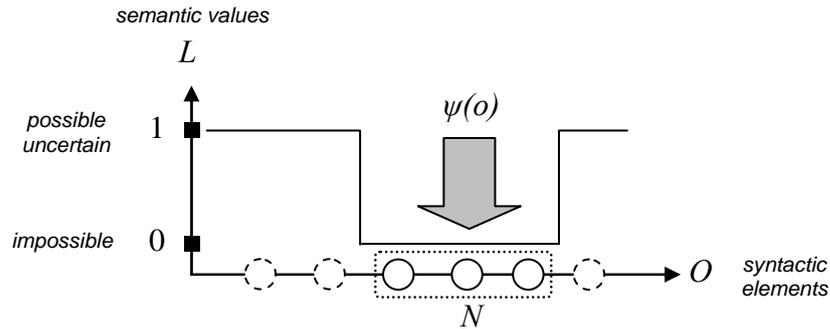

*Fig. 13. Possibilistic semantics.*

If data is normally represented in an extensional form by enumerating all semantically correct elements from set $M$ then constraints are normally represented in an intensional form by specifying characteristic properties of elements from $N$ which are known to be impossible. In other words, elements of subset $M$ are represented explicitly while elements of subset $N$ are represented implicitly.

One use of constraints consists in representing negative semantics by excluding elements which are known to be meaningless. For example, we might say that persons with age 200 years or higher are impossible. This information is stored in the database and then is used for consistency checks. The state is considered consistent if intersection of data with constraints is empty: $M \cap N = \emptyset$. In other words, the database does not have any positive claims about what is known to be impossible. Such constraints can be viewed as static because they are permanent part of the database, i.e., the database consists of two parts: positive $M$ and negative $N$.

Another use of constraints consists in selecting part of data stored in the database. Such constraints are dynamic because they are stored in queries and describe what is interesting to a user or application. For example, we might want to retrieve only data on persons with the age 30 years or older. Such constraints actually say that all younger persons have to be effectively removed and only persons satisfying this criterion are considered possible. In such a use constraints are imposed on the current data and their intersection $M \cap N$ is returned to the user.

In the two-level model an elementary constraint is imposed on one concept by excluding some its items which are declared impossible (while all the rest of the items from this concept are assumed to be possible). Constraints on concept $C$ are represented as a possibility distribution $\psi_C : C \rightarrow L$ which assigns some semantics value $\psi_C(e)$ to each element $e \in C$ and hence an element is either false or true. Elementary constraints cannot be overwritten, i.e., if an element is excluded then it cannot be made possible by imposing new constraints (otherwise we would get non-monotonic logic). Notice that elementary constraints use only this concept for selecting possible elements, i.e., we cannot use items from other concepts. For example, if concept Products has a sub-concept Manufacturer then we cannot select a subset of products produced by some manufacturer using only elementary constraints. Thus elementary constraints use only internal properties of items (represented and passed by value like references in CoP). For example, we might select an interval of ages because ages are represented by numbers.

Elementary constraints allow us to select a subset of items from individual concepts but the question is how we can propagate these constraints through the whole model. The thing is that normally removing an element from a concept affects also other concepts in this model, i.e., imposing constraints on one concept results in getting new constraints in some other concept(s). The simplest case of such constraint propagation arises when we interpret the ordered structure of items in terms of attribute values, i.e., a super-item is an attribute value for its sub-items (see section xx). Normally we want to select a subset of objects (from one concept) by specifying a subset of its properties (from the super-concept). If $C$ is a concept, $d$ is one of its direct dimensions with the domain in super-concept $D$ and elements of $D$ are constrained by the distribution $\psi_D$ then this constraint is propagated down to $C$ by producing distribution $\psi_C$ according to the following rule:

$$\psi_D(x) = 0 \;\wedge\; e.d = x \;\Rightarrow\; \psi_C(e) = 0$$

This definition means that if an item (attribute value) is prohibited by some constraint then the object it characterizes is also (automatically) prohibited. So an item is possible if only its super-items are still possible:

$$M = \{e \in C \mid \psi_D(e.d) = 1\}$$

The above formulated principle allows us to specify impossible values and then the objects characterized by them (along the specified dimension) will be automatically excluded from the set of possible items. In other words, constraints imposed on a concept are automatically propagated to its sub-concept(s). Yet here we have some asymmetry. The thing is that formally speaking we do not have a separation on values and objects and hence the propagated new constraint can itself propagate further down. In this case we simply assume that the excluded objects play the role of attribute values for the sub-items. The general principle of downward constraint propagation is formulated as follows:

> [Downward constraint propagation] An item is prohibited if at least one of its super-items is prohibited.

Thus in the general case constraints are propagated down to bottom concept recursively using the principle that if an item is assigned 0 (null) by some constraint (directly or indirectly) then its sub-items along selected dimension(s) are also assigned 0.

Obviously, this principle means that if an item is excluded then its de-projection of all ranks along the specified dimension is also excluded (Fig. 14):

$$\psi_D(x) = 0 \;\Rightarrow\; \psi_F(y) = 0, \forall y \in I = x \leftarrow f \leftarrow F, \forall \operatorname{rank}(f)$$

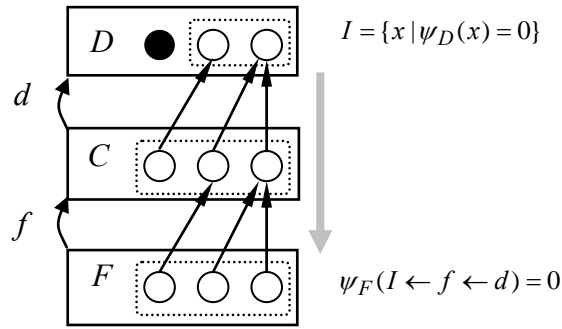

*Fig. 14. Downward constraint propagation.*

We said already that imposing constraints means assigning 0 (null) as a semantic value to some items so that these item look like non-existing. This entails impossibility to use them as super-items (from sub-concepts). However, if some item from a sub-concept already used them, i.e., $\psi_C(e) = 0$ and $i = \langle \ldots, d : e, \ldots \rangle$, then the definition of this sub-item changes. Namely, instead of the prohibited item we write 0 (null) which essentially means that this super-item is absent, $i = \langle \ldots, d : null, \ldots \rangle$. At this moment we get an alternative concerning what to do with the modified sub-item: (1) if it can exist with null as a super-item then it remains possible, i.e., $\psi_D(i) = 1$, (2) if it cannot exist with null as a super-item then it is marked impossible. $\psi_D(i) = 0$. In the latter case this sub-item is prohibited as a result of prohibiting its super-item and hence then the procedure proceeds recursively to the next sub-item. Earlier we assumed that the second alternative is true.

Above we considered how constraints imposed on a concept are propagated down to sub-concepts. However, constraints can also propagate in the opposite direction from a concept to its super-concepts. Informally, the idea is that if an object is removed from consideration then we are also not interested in having its attribute values. For example, if we select a subset of products then we are interested in seeing only their manufactures rather than all possible manufacturers. Yet a value is excluded only if it is not used by *any* selected object, i.e., a value is still possible and visible in the selection if at least one object uses it in one of its attributes. In the general case the upward propagation principle is formulated as follows:

[Upward constraint propagation] An item is prohibited if *all* its sub-items are prohibited.

Formally this means that if an item is possible then all its super-items are also possible:

$$\psi_F(x) = 1 \wedge x.d = e \Rightarrow \psi_D(e) = 1$$

Obviously, this means that only the projection of possible items is possible while all other items are impossible (Fig. 15):

$$\psi_F(x) = 1 \Rightarrow \psi_D(y) = 1, \forall y \in I = x \rightarrow f, \forall \text{rank}(f)$$

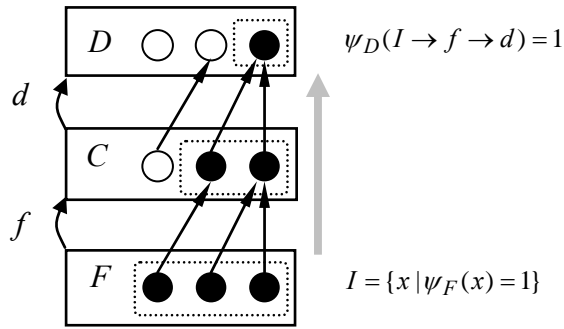

*Fig. 15. Upward constraint propagation.*

Upward propagation is a propagation of possibility while downward propagation is a propagation of impossibility (prohibition). In other words, if an item is marked impossible then all its sub-items are marked impossible (assuming that initially all are possible) and if an item is marked possible then all its super-items are marked possible (assuming that initially all are impossible). In geometric terms, if a coordinate is removed then all points having it are also removed and if a point is placed then all its coordinates must be also present.

### 4.4 Dependencies and Inference

Let us assume that variable $y \in Y$ depends on $x \in X$. In the case of no dependence variable $y$ could take any value from $Y$ without any restrictions. However, if $y$ depends on $x$ then selecting a subset of values in $X$ means restricting possible values in $Y$. The presence of dependence can be expressed in terms of three sets: $X$, $Y$ and their Cartesian product $X \times Y$ which consists of all combinations of values from $X$ and $Y$. The dependence can be represented as a subset of combinations of values, $Z \subseteq X \times Y$. If we restrict values in $X$ by selecting its subset $X' \subseteq X$ then the set of possible points in $Z$ is also restricted and we get its subset $Z' \subseteq Z$. The subset of points then is projected to set $Y$ and restricts it so that we get a subset of values from $Y' \subseteq Y$.

In two-level model dependencies are treated analogously to this interpretation. Concepts are treated as sets of values and selecting some subset is performed by imposing and propagating constraints. The main difference is that concepts in our approach are ordered and hence their relative position determines their role. In particular, a common sub-concept plays a role of dependence with respect to its super-concepts. For example (Fig. 16), if concept $Z$ has two super-concepts $X$ and $Y$ then its semantics (its set of items) represents a subset of all possible combinations of items from *X* and *Y*. By imposing some constraints on concept *X* we can select its subset *X'*. This subset is then propagated down to concept *Z* where we select subset *Z'*. Finally, subset *Z'* is propagated up to concept *Y* where we get subset of items *Y'*. Thus having a common sub-concept allows us to propagate constraints from *X* to *Y*. Obviously, this procedure is symmetric and hence we say that concepts *X* and *Y* are dependent.

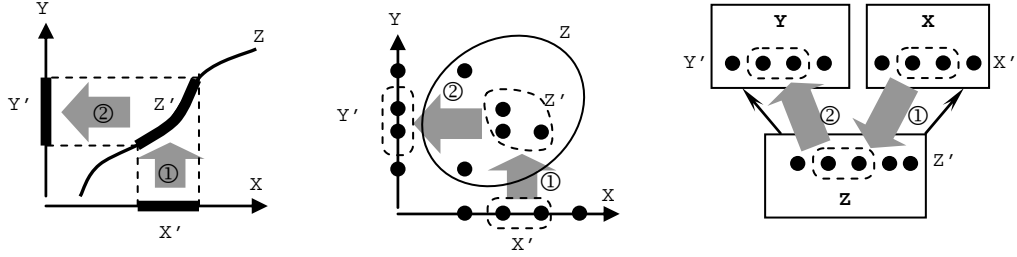

*Fig. 16. Dependencies and inference.*

The procedure of propagating constraints from one concept to a dependent concept is referred to as *inference*. In the general case this procedure depends on many factors. For example, it depends on the nature of the set of semantic values *L*. If it is an interval [0,1] with operations min and max then we can define fuzzy inference. If we use probabilistic operations then we get probabilistic inference where probabilities over the set *X* are propagated to probabilities over *Y* taking into account probabilities of *Z* (such as factor analysis). For the inference procedure it is important also what kind of data/knowledge representation is chosen. For example, we might use rules to describe dependence of *Y* from *X* or analytical formula like in neural network.

In the described approach we assume that data is represented explicitly, i.e., each item is represented by its own individual and unique reference. Then the procedure of inference consists of two major steps (Fig. 16):

1. De-projecting source constraints over selected sub-dimensions down to bottom concept
2. Projecting the obtained semantics of bottom concept up to the target concept

Input constraints can be imposed independently on several concepts $X_1, X_2, \ldots, X_n$. Independence means that they are propagated down to bottom concept and then the final result is intersection of the input de-projections:

$$Z' = (X_1 \leftarrow x_1 \leftarrow Z) \cap (X_2 \leftarrow x_2 \leftarrow Z) \cap \ldots \cap (X_n \leftarrow x_n \leftarrow Z)$$

Notice that many propagation dimensions can be specified for each one input concept and each such propagation dimension makes its own independent contribution to the final constraint imposed on bottom concept *Z*.

The output semantics is produced on the second step of the inference procedure by projecting the constrained bottom concept semantics *Z'* up to the output concept *Y* along the chosen dimension *y*:

$$Y' = Z' \rightarrow z \rightarrow Y$$

# 5  Uses of the Model

## 5.1  Query Language

In this section we describe a simple query language which is mainly intended to demonstrate possibilities of the concept-oriented data model. It is important that it is not a full featured query language but rather a set of language constructs which can be incorporated in other languages or developed further. We chose SQL-like syntax for this language because SQL currently dominates among query languages and is therefore easy to comprehend. Yet, in some cases we propose alternative keywords and syntactic constructs which seem more appropriate for our model and emphasize its difference from the existing approaches.

In two-level model there are two types of elements: individual data items and collections of items. Both types of elements are represented by their references which can be stored in variables. Concepts in the database are actually collections which are globally visible and accessible via their name. However we can also create dynamic collections within the current query scope. Reference to a collection can be returned by queries and then it can be stored in a variable. Such variable is declared as having type `Collection`. For example, if we want to store a collection of items returned from a query then we write it as follows:

```
Collection myCollection = SELECT * FROM Employees;
```

Returned collections can be also used anonymously, for example, by passing them as parameters or applying to them an operator directly:

```
AVERAGE( SELECT age FROM Employees );
SELECT * FROM (SELECT * FROM Employees) WHERE age > 30;
```

Item identity (reference) can be stored in variables or used anonymously. We can apply dot operation to an item variable followed by dimension name in order to access content stored in this item entity, i.e., a reference to the corresponding super-item:

```
Item mySuperItem = myItem.myDimension;
```

Collections and items have some structure which is described by its dimension names and super-item types. If we need only one dimension then it can be selected using the same dot operation applied to a collection:

```
Collection collectionOfSuperItems = myCollection.myDimension;
```

Notice that the new collection has the same number of element, i.e., it is not projection. If it is necessary to select several dimensions then they can be specified in angle brackets:

```
Collection newCollection = myCollection.<dim2, dim4, dim6>;
Item newItem = myItem.<dim2, dim4, dim6>;
```

In strong typed systems it is necessary to specify a type of an element referenced by a variable. In our case it is desirable to be able to restrict type of collections and items rather than marking then as `Collection` and `Item`, respectively. We will assume that item type is specified by the collection name. For example, if we have collection (say, concept) `Employees` then it can be used as a type of items and these items will be automatically restricted to the structure of this collection:

```
Employees manager = getManager(person);
```

Here we pass a person reference to query `getManager` which returns a reference to his/her manager which is then stored in a typed variable. Notice again that here type `Employee` is used to restrict the structure of the variable.

Type of collections is specified in angle brackets as follows:

```
Collection<Employees> persons;
```

Here we declared a variable which references a collection of items where items are of type `Employees`.

A collection is not an isolated set of items. Rather, it lives within some structure of super-items and possible sub-items (if its elements are referenced from some other collection). Hence we can apply operations of projection and de-projection to collections as well as individual items which are denoted by right arrow and left arrow followed by dimension name, respectively. For example, if we have a collection of products then a set of their manufactures can be obtained as follows:

```
Collection companies = products -> manufacturer -> Companies;
```

Here we use concept `Companies` for projecting. In the general case it can be any collection.

If we have a manufacturer then the set of its products can be obtained using de-projection:

```
Collection products = companies <- manufacturer <- Products;
```

Instead of concept `Products` we might use any collection with a subset of products. Projection and de-projection can be combined by producing a so called zigzag query.

In addition to restricting dimensions of a collection, frequently we would like to restrict its items by selecting only those satisfying certain criteria. In our language these criteria are separated from the source collection by bar symbol, i.e., we write the input collection then bar symbol and finally a set of conditions each items in the output collection has to satisfy. For example, if we want to select only persons with the age 30 then it can be done as follows:

```
Collection p = (Employees | age == 30);
```

Here condition uses dimension names of the input collection. Actually, such a form is a shortcut of the general case described below. In particular, we omitted instance variable for the input collection that can be then used in condition part:

```
Collection p = (Employees emp | emp.age == 30);
```

Condition may have any complexity but normally it uses the instance variable and dimensions names. Projection and de-projection operations are also possible. For example, we might select only persons who participate in more than 3 projects:

```
Collection p = (Employees e |
    COUNT( e <- employee <- ProjEmp ) > 3);
```

Here each current person from concept `Employees` is de-projected down to concept `ProjEmp` (a link between projects and employees) and we evaluate the size of this de-projection. Since it is a normal collection we restrict the number of its elements using aggregation function `COUNT`.

In the general case a query consists of the following parts:

> `FROM` – input collections (the structure of space)
>
> `SELECT` – output structure (horizontal restrictions)
>
> `WHERE` – semantic constraints (vertical restrictions)

First of all any query is supposed to get some input collections which are specified using their references such as concept names or references to collections. Input collections are specified in the `FROM` clause where they are separated by comma. For each collection an instance variable is also specified. For example, if our query processes two concepts `Projects` and `Personnel` then it is written as follows:

```
FROM ( Projects project, Personnel personnel )
```

The result of `FROM` clause can be thought of as a description of new space which consists of all combinations of the input items (i.e., the Cartesian product of input collections). In the last example this clause says that we are going to consider all combinations of projects and persons. An alternative keyword for this clause could be `FORALL` which means that the query will produce its result from the set of all combinations of items in the specified collections.

A query with a single `FROM` clause will actually return a new collection which consists of all combinations of its input items and has all dimensions from the input collections Notice that this result is a normal collection which can be used in other queries or stored in a variable.

`SELECT` clause is intended to restrict the structure of the returned collection by specifying only the input dimensions we want to have in the output collection. For example, in the following query for all combinations of projects and persons we return only two dimensions:

```
FROM ( Projects project, Personnel personnel )
SELECT ( project.name, person.name )
```

An alternative name for the same construct could be `RETURN` which emphasizes that the specified values have to be returned in the output collection. Of course, we can provide computed values in this clause including projections and de-projections.

The main purpose of any query consists in selecting a subset of items from the input multi-dimensional space described in `FROM` clause. It can be done in `WHERE` clause which provides a complex condition each input item has to specify. In other words, the output collection will consist of only input items which satisfy the provided condition. The criterion is a logical formula consisting of elementary conditions and logical connectives. For example, we might restrict items as follows:

```
FROM ( Projects project, Personnel personnel )
```

```
SELECT ( project.name, person.name )
WHERE ( person.age > 30 AND project.budget < 20 )
```

Notice that we use instance variable declared in `FROM` clause to access other items in the whole model. In other words, all items are living within one global structure and we can use this structure for access.

`WHERE` clause allows us to provide a condition in a declarative form and then the query will return all input items which satisfy it. In some cases conditions could be provided in an imperative form as a procedure which is written as a query body. For example, the previous query could be written as follows:

```
FORALL ( Projects project, Personnel personnel ) {
  IF ( person.age > 30 AND project.budget < 20 )
    RETURN (project.name, person.name );
}
```

Here the declarative and imperative approaches are syntactically equivalent (essentially we changed only keywords and their order). Notice that in this example the order of computations is not specified. The imperative form (with query body) is more flexible in writing very complex queries especially with manual computations of intermediate elements. Such queries use full power of imperative approach but have much less possibility for optimization and loose elegance and simplicity of declarative queries.

## 5.2 Multi-Valued Properties

By property we normally mean some element associated with this element. However, frequently when describing a problem domain it is desirable to associate a collection of other elements as a characteristic of this element. Such collections can still be considered normal properties especially if they are named. For example, an order could be characterized by a set of order parts or a customer could be characterized by a set of made orders. One traditional solution to this problem consists in introducing a new dedicated mechanism which allows us to have multiple-valued properties. Such values can be viewed as arrays or lists and normally are used for storing small collections. Although this approach could be convenient in some cases, it has one drawback: it is not clear how to formally interpret such an extension.

Formally, in the concept-oriented model we have only ordering of elements while other mechanisms are only interpretations of the order. In particular, the ordering of elements can be used to model multiple-valued properties. The main idea is that such properties are associated with sub-items of elements. If we have an item then its super-items are interpreted as single-valued properties while its sub-items are interpreted as multiple-values of some property. Obviously, multiple-valued properties are de-projections of this item along some sub-dimension. For example, an order is characterized by a set of order parts which are obtained by de-projecting to the corresponding sub-concept.

The only thing that lacks here is the name for properties. Indeed, if for single-valued properties we can use dimension names then for multiple-valued properties such an approach does not work because sub-dimensions are not always unique. In order to overcome this difficulty a derived property can be defined for any concept. A derived property is actually an arbitrary named query which returns some result set. For example, we could define a multiple-valued property of concept `Employees` which returns a set of its orders:

```
Employee::orders() {
  RETURN this <- employee <- Orders;
}
```

Once this property is defined it can be applied to employee items as if it were normal property.

In the general case derived properties may take input parameters. For example, we might define a property which returns a set of orders having some category:

```
Employee::orders(Category c) {
  RETURN this <- employee <- (Orders o | o.dish.category == c);
}
```

Derived properties could have any definition which is allowed for normal queries.

## 5.3 Grouping and Aggregation

The mechanism of grouping and aggregation is highly important for data modelling. It has many interpretations and possible applications but the general idea is that focus is shifted to manipulating groups of data elements such as subsets of records. It is important that a group is considered an entity with its own properties and characteristics. For example, in such a group-oriented approach departments are characterized not only by name and location but also by its employees and projects, i.e., one department has a group of employees working in it and a group of projected associated with it (possibly indirectly via employees). In this approach two types of elements are supposed to exist: (i) groups and (ii) group constituents. For example, we might consider departments as groups and employees as their constituents. Or we might consider employees as groups and projects as their constituents.

In relational model grouping is based on specifying a characteristic property of group constituents which has to be the same for one group. Thus group elements do not participate in such an approach explicitly. For example, we could group all employees depending on their department property, i.e., all employees having one department id produce one group. Notice however that departments do not participate in this operation explicitly. Yet departments (groups) are precisely what will be produced in the result set while employees (group constituents) are used only to produce local groups and compute some (aggregated) properties. Thus in RM one relation can be grouped in many different ways depending on the characteristic property specified in the query.

In the concept-oriented model grouping is not a separate dedicated mechanism so we do not need special operations or additional support from the model. Grouping is a natural consequence of the model structure. Namely, it is a consequence of the third interpretation of the ordering relation (see section xx) which says that an element is a logical collection or group of its sub-elements. Thus groups are not specified explicitly in queries but exist permanently in the database. For example, if a department has a number of employees as its sub-items then these employee items are known to constitute a logical group associated with this department. Obviously, a group associated with this element can be retrieved as its de-projection along some sub-dimensions down to some sub-concept. In this approach one element is characterized by many different groups from its sub-concepts, i.e., group has one type while constituents have other types.

Below we consider how this approach can be used for data retrieval using example shown in Fig. 17. Let us assume that employees (concept `Employees`) are ordering dinner (concept `Orders`) which is characterized by date (concept `Dates`). Each dinner consists of several dishes (concept `Dishes`) characterized by dish type (concept `Categories`) and price. The many-to-many relationships between ordered dinner and dishes is stored in concept `OrderParts` which is bottom concept in this model containing the most specific data items.

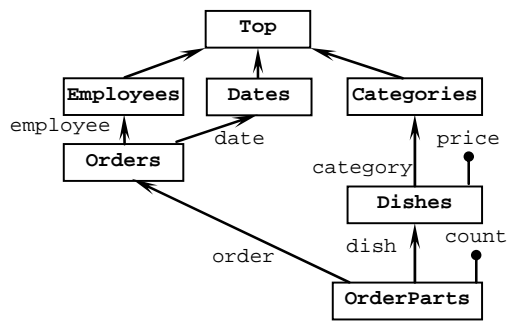

*Fig. 17. An example of grouping and aggregation.*

Assume we need to get a list of persons who ate 'pizza' more than 100 times during 2006. From this problem formulation we see that we need to get persons as an output so we start our query concept `Employees` which is the only input concept for the query:

```
FROM Employees e WHERE ... SELECT ...
```

What persons to return is already the second issue in this approach which is deliberately separated in the query syntax (in SQL we would write all the participating table names in `FROM` clause). For each person in this model there is a group of orders and each order has a group of order parts. We need to choose persons depending on what concretely they ordered during 2006. In our query we can describe this group by using de-projection:

```
Collection group = e
   <- employee <- Orders
   <- order <- OrderParts;
```

However, we do not need all order parts for the current employee – it is necessary to choose only those made in 2006 and belonging to the specified category. So we restrict items in the two concepts as follows:

```
Collection group = e
   <- employee <- (Orders | date == '2006')
   <- order <- (OrderParts | dish.category == 'pizza');
```

Alternatively, we might write it as follows:

```
Collection group = e <- employee <- order <- (OrderParts |
order.date == '2006' AND dish.category == 'pizza');
```

In both cases variable `group` will contain a group of employees who ordered 'pizza' in 2006. Now for getting the final result we need to simply choose employees for which this variable has more than 100 items:

```
WHERE COUNT( group ) > 100
```

Finally we need only to pack these fragments in one query:

```
FROM Employee e
WHERE COUNT(
    e <- employee <- order <- (OrderParts
       | order.date == '2006' AND dish.category == 'pizza')
   ) > 100
SELECT e.name
```

If we need to compute more properties of the employees using the obtained group of order parts then its reference could be stored in a local variable within query body. For example, if in addition to person name we need to return the total price then it can be done as follows:

```
FROM Employee e
{
   Collection group = e <- employee <- order <- (OrderParts
      | order.date == '2006' AND dish.category == 'pizza');
   double total = SUM( group.<count * dish.price> );
}
WHERE COUNT( group ) > 100 AND total < 200
SELECT e.name, total
```

Here the intermediate values are computed in the body. In particular, we compute the group of order parts satisfying our conditions and then sum up the price paid for them. Then we choose only persons who ate more than 100 pizzas and paid less than 200 EUR for that. The output includes name of the person and the total price paid. The same could be done without using query body but with much less clarity. An equivalent but more imperative form of the same query is as follows:

```
FROM Employee e
{
   Collection group = e <- employee <- order <- (OrderParts
      | order.date == '2006' AND dish.category == 'pizza');
   double total = SUM( group.<count * dish.price> );
   IF( COUNT( group ) > 100 AND total < 200 )
   THEN RETURN( e.name, total );
}
```

An alternative more verbose form of the same query can be written as follows:

```
FROM Employee e
{
```

```
    Collection group =
      FROM OrderParts op
      WHERE
        op.order.date == '2006' AND
        op.dish.category == 'pizza'
      SELECT op.count * op.dish.price
}
WHERE COUNT( group ) > 100 AND SUM( group ) < 200
SELECT e.name, SUM( group ) AS total
```

Here instead of using de-projection operation, we manually build de-projection by selecting items from `OrderParts` which reference the current employee item from `Employees`.

All these queries have the following important distinguishing features:

- We do not use joins for getting related items. Such queries are easier to write and clearer for understanding.
- We use projection and de-projection along dimensions.
- Groups are collections built within an external query.
- Aggregated properties are normal properties computed from internal collection.

Grouping in SQL is done as follows. First, we need to select elements of groups satisfying their conditions (`WHERE` clause in SQL). Then the selected elements are grouped (`GROUP BY` clause in SQL) and their aggregated properties are computed. And finally only groups satisfying their properties are selected (`HAVING` clause in SQL). Notice that here we have two kinds of restrictions: those imposed on elements of groups via `WHERE` clause and those imposed on groups themselves via `HAVING` clause. Thus it is assumed that only one group is produced from the records.

Let us now suppose that we need to take more aggregated properties into account which are computed on different groups. In the above example we selected employees depending on how many pizzas they ate in 2006. What if among them we need to select only employees who made at least 10 orders in 2007? Notice that for each person we need to create and analyze two groups: a collection of order parts (with pizza) made in 2006 and a collection of orders of any kind made in 2007. This query is written very simply by selecting the persons in the outer query depending on the properties of groups created inside this query:

```
FROM Employee e
{
  Collection group =
    e <- employee <- order <- (OrderParts
      | order.date == '2006' AND dish.category == 'pizza');
  Collection group2 =
    e <- employee <- (Orders | date == '2007');
}
WHERE COUNT( group ) > 100 AND COUNT( group2 ) > 10
SELECT e.name
```

We might also easily add more groups created over employees such as a collection of different dish categories. What is common to all these queries is that in the outer query we select group items (persons) using either their existing properties like age or properties computed inside the query body. In particular, such computed properties could be produced by creating collections of items using de-projection or more complex queries.

One general problem in dealing with groups of data items consists in mutual influence of different restrictions. This problem is especially complicated if restrictions belonging to logically different elements of the query are imposed on the same items. For example, let us assume that we want to select employees who ate such dishes in 2006 that were eaten by more than 10 different employees in 2007. Here we need to select employee items but we see also that these very items influence dish categories which in turn are taken into account when selecting employees. Another complication is that we have two incompatible restrictions on dates of orders which also have to be somehow arranged.

In our approach writing such a query is not much more difficult than in the previous example. The general idea is that we start from what we want to get and produce the necessary properties that are used to restrict the selection. In this example the selection of employees depends also on employees, i.e., each employee is a group of other employees who ate the same dishes. This can be written as follows:

```
FROM Employee e
{
  Collection group =
    e <- employee <- order
      <- (OrderParts | order.date == '2006')
      -> dish -> category
      <- category <- dish
      <- (OrderParts | order.date == '2007')
      -> order -> employee;
}
WHERE COUNT( group ) > 10
SELECT e.name
```

Notice that in this query we go through concept `OrderParts` two times with different date restrictions. We can also easily add more restrictions to other intermediate items including additional aggregated properties. For example, instead of selecting orders belong to a concrete year we can select those orders that have a year characterized by a high number of orders:

```
(OrderParts | COUNT(order.date <- date <- Orders) > 1000)
```

This fragment selects belonging to a year with high number of orders (more than 1000).

## 5.4 Multi-Dimensional Analysis and OLAP

The mechanisms based on projection and de-projection operations described in the previous sections are inherently one-dimensional because we always move along one path in the concept graph. We can change the path and its direction or we can provide criteria the intermediate elements have to satisfy. However, this method does not allow us to move along several paths and select elements taking into account several project and de-projection operations executed in parallel rather than sequentially.

In practice such a one-dimensional approach is frequently very restricted because we need to view elements as belonging to several groups simultaneously. For example, we might want to view all the sales as belonging to some country *and* to some product category. In this case each concrete sale belongs to one cell in the two-dimensional space where one dimension consists of all countries and the second dimension consists of all product categories. On the other hand, each cell from the two-dimensional space, i.e., each combination of one country and one product category, is a group of sale facts. Such type of analysis where elements are viewed simultaneously along several dimensions is studied in online analytical processing (OLAP). In the concept-oriented model it can be performed using a procedure consisting of the following steps:

1. Choose a *fact concepts* consisting of items that will be grouped along several dimensions and impose constraints on its elements
2. Choose several *dimension paths* of the fact concept along which its elements will be grouped
3. Choose one *level concept* along each dimension path and impose constraints on them
4. Build a *multi-dimensional cube* as the Cartesian product of all the level concepts
5. Group elements of the fact concept over elements of the multi-dimensional cube
6. Choose a measure of the fact concept and compute its aggregated value for each group

The result of such an analysis will be one aggregated measure property associated with one cell of the multi-dimensional cube.

Let us assume that we need to analyse how our company sales are distributed in the space of customers and products. Each individual sale is stored as an item of concept `OrderParts` (Fig. 18). Items of this concept will be grouped over two dimensions and some property of this item will be then aggregated. Hence we choose `OrderParts` as a fact concept in our analysis.

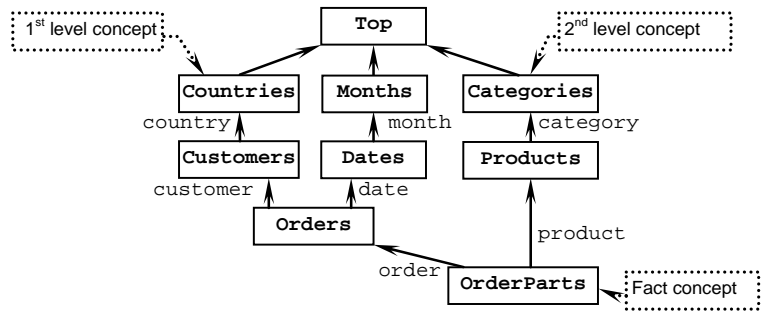

*Fig. 18. An example of multi-dimensional analysis.*

The first dimension path describes customers: `OrderParts->order->customer->country`. Thus any fact item (one sale) is characterized by an order, a customer or a country, depending on the level of details we choose on the next step. The second dimension path describes products: `OrderParts->product->category`. Now any fact (one sale) is characterized by two values: one from the customer dimension path and the second from the product dimension path.

Each dimension path chosen on the previous step consists of several concepts which differ in their level of details. For analysis we need to choose one level of details for each dimension path. For example, we might start from the lowest level of details by choosing `Countries` as a characteristic of customers and `Category` as a characteristic of products. If later we need to see more details then we can down to a sub-concept (drill down in OLAP terms). After choosing these level concepts, each fact item belongs simultaneously to a pair of level items. In other words, one sale belongs to one country *and* one category as a group.

Now let us consider how fact items are grouped over elements of multi-dimensional space. First of all we need to produce a multi-dimensional space from the chosen level concepts. This can be done using `FROM` clause where level concepts are parameters:

```
FROM (Countries country, Categories category) ...
```

The result of this query is a collection consisting of all combinations of countries and categories. If we do not need the whole space then its size can be restricted by imposing constraints in `WHERE` clause. For example, if we want to consider only countries with at least one customer then it is written as follows:

```
FROM (Countries cntr, Categories ctgr) ...
WHERE ( cntr <- country <- Customers > 0 )
```

Here we de-project this country and compare the size of the group of obtained customers with zero. We used a shortcut which in full form is written using aggregation function:

```
WHERE ( SIZE(cntr <- country <- Customers) > 0 )
```

Now it is necessary to group fact items over the cells of the two-dimensional space. This can be done manually by selecting fact items for each current pair country and category:

```
Collection grp =
    FROM OrderParts op
    WHERE op -> order -> customer -> country = cntr
        AND op -> product -> category = ctgr
```

Here we actually use a definition of multi-dimensional de-projection by selecting only items which are projected along the both dimensions on the current pair of country and category. If multi-dimensional de-projection is supported in the query language then it could be written in the following shorter form:

```
Collection grp = [
        cntr <- country <- customer <- order <- OrderParts
        AND ctgr <- category <- product <- OrderParts
    ]
```

This syntactic format (using square brackets) takes two or more de-projections starting from different super-concepts (level concepts) and ending in one sub-concept (fact concept). The result of each group (multi-dimensional de-projection) is an intersection of all one-dimensional de-projections constituting this query.

When building a group of fact items for each cell in the two-dimensional cube we can restrict its elements by imposing some constraints. For example, we might select sale facts only for some year:

```
Collection grp =
    FROM OrderParts op
    WHERE op -> order -> customer -> country = cntr
        AND op -> product -> category = ctgr
        AND op.date = 2007
```

Now we have two-dimensional space for each cell of which a group of facts is built. However, we would like to see some aggregated property of this group rather than a number of its elements as an output. In this case it is necessary to choose an aggregated property as a measure. For example, we might sum up the price paid for the orders within one group:

```
double total = SUM ( grp.price )
```

It is possible to select more measures the number of orders in the group of order parts:

```
integer cnt = COUNT ( grp -> order )
```

These measures are then included in the query output via SELECT clause:

```
SELECT cntr.code, ctgr.id, total, cnt
```

The whole query is written as follows:

```
FROM (Countries cntr, Categories ctgr)
WHERE ( cntr <- country <- Customers > 0 )
    Collection grp =
        FROM OrderParts op
        WHERE op -> order -> customer -> country = cntr
            AND op -> product -> category = ctgr
            AND op.date = 2007
    double total = SUM ( grp.price )
    integer cnt = COUNT ( grp -> order )
SELECT cntr.code, ctgr.id, total, cnt
```

In the case we need more detailed analysis it is possible to choose other level concepts along the dimension paths. If we move down in the graph and choose a sub-concept then this operation is normally called drill down. If we move up and choose a super-concept along this dimension path then this operation is called roll up.

## 6   Related Work

CoM allow for modelling the hierarchical structure of its elements which is called physical structure. In this sense it is analogous to the classical *hierarchical data model* where elements are related to each other using one-to-many parent-child relationship. The main difference of CoM is that its hierarchical structure is intended for identity modelling, i.e., we use the hierarchy for modelling how data items are represented and accessed. In particular, using the inclusion hierarchy we can define the structure of complex references used to uniquely identify elements of the model. However, this structure is not used to represent data semantics. Thus CoM allows us to use all the numerous advantages of having a hierarchy at the same time providing other techniques for modelling data semantics.

In CoM, data elements can reference each other within logical structure. Such connections can be viewed as a network of data elements and in this sense CoM is similar to the *network data model*. The main difference of CoM is that its connections represent ordering relation among elements rather than an arbitrary graph. Another difference is that elements in CoM exist in a hierarchy as described above. Thus CoM provides two orthogonal structures: physical hierarchy and logical ordering. Physical hierarchy describes how elements are represented and accessed while ordering represents data semantics. Thus CoM combines features inherent to both hierarchical and network models. The navigational features of CoM make it similar to the functional data model (FDM) [Shi81, Gra99, Gra04].

If we ignore the order of elements and consider them as members of sets without hierarchies then CoM is analogous to the relational data model (RM) [Cod70]. In particular, CoM allows for relational operations over its elements if we interpret concepts as relations. For example, it is possible to apply join operations to concepts by producing new sets of items. The main difference of CoM in comparison with the relational model is that CoM relies on ordering relation when describing model syntax and semantics while the relational model manipulates sets only. Thus we reach a synergy effect in the concept-oriented model by taking all the best from the classical models and implementing different mechanisms using very simple and intuitive principles.

One important difference between CoM and RM is the treatment of data types. RM divides the whole structure into data types and relations (which then can be defined using different domains). Thus on one hand we can develop a structure of data types, for example, using inheritance as it is done in OOP. An on the other hand we develop a structure of relations. However, this division is not symmetric because RM concentrates mainly on relations while data types are in great extend an auxiliary facility. In CoM, such a division does not exist. However, we introduce duality between identity modelling performed via physical structure and entity modelling expressed in logical structure. Informally, if projected in terminology of RM, physical structure in CoM can be used to model data types in RM while logical structure in CoM can be used to model relations in RM. This division is symmetric and both structures are equally important for the concept-oriented model. In other words, identity modelling is integral part of CoM where we define the structure of our data space using inclusion relation and a hierarchy of data items. Logical structure in CoM is used to define data semantics which is essentially absent in RM.

One problem of RM is that it is flat model because all relations have equal rights, i.e., they are members of one collection. However, most problem domains have normally rather complex structure which has to be represented via a flat set of relations. This problem can be viewed as another side of the advantages obtained from the flat system of relations. Historically, the hierarchical model and the network model provided rather good means for describing a structure however these methods were too restrictive or too complex. RM switched to a flat system of relations and gained simplicity however such an approach did not supported directly hierarchies and other structures which are widely used in practice. As a response to this demand various modifications to RM have been developed with the purpose to support structured designs. One of the most known extensions consists in introducing relation-valued attributes (rather than only scalar-values). This extension looks very natural because in most designs the modeller deals with properties which take a value which is itself a set. Moreover, the members of this set are complex elements which have their own properties. For example, a customer could be characterized by a set of orders and then each order is characterized by a set of product items. Thus the nested-relation approach aims to reflect such a problem domain design directly in the model rather than to model this situation manually using numerous auxiliary relations and complex queries. This model provides a generalization of RM by hiding the underlying structure of relations which are used to specify queries. The main goal consists in working with relations obtained from queries directly as attribute values rather than as selections produced by complex queries involving many relations.

CoM provides a natural solution to this problem based on de-projection operation which allows for retrieving sets of items associated with this item. Thus CoM retains the original design where the problem domain is described using a number of concepts at the same time providing a very easy and natural way for working with multiple-valued attributes in a nested manner. For example, in nested relation model if a customer has many orders then it can be expressed as follows:

```
Customer(
    id: INTEGER,
    name: CHAR(64),
    orders: SET OF (id: INTEGER, date: DATE, price: MONEY ...)
)
```

Notice that here attribute `orders` is a relation-valued attribute defined via `SET OF`. Such a schema can be queries using an extended SQL which allows for predicates and operators applied to relation-valued attributes. For example, we could select all customers having orders with some concrete date:

```
SELECT id FROM Customer WHERE orders.date = '31.12.2007'
```

Notice that here we use dot notation to access attributes for the nested relational value. If we add an additional condition imposed on set-values then they can be interpreted ambiguously. For example, the following query has at least two natural interpretations:

```
SELECT id FROM Customer WHERE
    orders.date = '31.12.2007' AND
    orders.price = 100
```

The first interpretation will return customers who have among their orders at least one with the specified date and at least one with the specified price. The second interpretation will return customers who have among their orders at least one record with the specified date and the specified price. More complex queries have even more problems with their interpretations which can be overcome but significantly increase complexity of the whole approach. In particular, such difficulties appear for many-to-many relationships. Normally, it is necessary to choose an order among relations so that they become unequal. Different clauses in SQL queries correlate and are ambiguous.

CoM does not have such problems because we work directly with concepts and de-projection operation which avoid ambiguity. For example, the following query returns all customers who have at least one order with the specified date and at least one order with the specified price:

```
SELECT id FROM Customer c WHERE
    c <- customer <- ( Order o | o.price = 100 ) > 0 AND
    c <- customer <- ( Order o | o.date = '31.12.2007' ) > 0
```

The next query returns all customers having at least one order with the specified date and price in two fields:

```
SELECT id FROM Customer c WHERE
    c <- customer <- ( Order o |
        o.date = '31.12.2007' AND o.price = 100 ) > 0
```

Another problem of RM is the absence of canonical semantics. As a consequence it is possible to produce very different designs which are equivalent but cannot be formally compared. Thus RM provides very limited facilities for analysing different syntactic designs and data semantics (it is aimed mainly at providing algebraic operations for producing new relations from existing ones). The mechanism of normalization provided by RM proposes several standard forms which have different properties but it does not allow us to say what is the meaning of the data and how different databases can be compared. One solution to this problem consists in providing a kind of universal representation which would hide all possible table designs and could be used as a high level of data view. This view would allow representing databases in a universal canonical form and compare their semantics.

As a response to this demand a special modification of the relational model was developed which is called the universal relation model (URM) [Ken81, Fag82, Mai84]. This model allows viewing the database as if it were composed of a single relation while all other information embedded into attributes. Having one relation at high level allows us to avoid ambiguity when interpreting different databases and therefore such a relation is called universal. In URM all relations are assumed to be projections of a single relation. However, this direction did not result in an acceptable solution because some things become simpler while others become more complex. One reason is that an assumption of universal relation was shown to be incompatible with many aspects of the relational model within which it was being developed.

In CoM, bottom concept can be viewed as a universal relation because it contains all the dimensions and the most specific data items which can be used to access all the other information in the model. It possesses canonical semantics so that data from different models can be represented in one form which allows for comparisons. It makes it possible to define such more complex mechanisms as constraint propagation and inference.

Thinking in terms of attribute-values is very natural for describing elements however originally this approach has only one level while most problem domains are described using a hierarchy. Although the relational model can be used for describing very different structures including a hierarchy, it is desirable to have an approach which would allow doing this at the level of the model itself. In other words, since the world is described using hierarchies it is necessary to reflect the hierarchy in the model used to describe the world. There exist different approach to multidimensional modelling most of which are based on the notion of *dimension* and *data cube* [Li96, Agr97, Gys97, Ngu00, Tor03, Mal06]. Dimensions allow us to introduce levels of details while data cube is a collection of facts.

CoM can be viewed as a development of this direction because it also allows for treating data as existing in a multi-dimensional space. However, the main difference of CoM is that it is an integrated model rather than a mechanism for introducing dimensions into some existing (logical) model. In other words, CoM is independent of any other model (although can be implemented over some existing model) while multi-dimensional models normally are defined as a new conceptual level over some model. Dimensions in CoM are a primary element of the model which then determines all the other properties and mechanisms. Dimensions are associated with ordering relation which is a corner stone of the concept-oriented approach. In CoM, we start from ordering elements using dimensions and then define their semantics and other properties. The traditional approach consists in establishing a new abstract level over some other model (normally relational model) so that the basic mechanisms are taken from the relational model while multi-dimensionality and other mechanisms are added as an additional feature. Although the underlying level (RM) is used only as physical storage it still influences how the multi-dimensional level is organized.

One of main features of CoM is that it is based on the theory of ordered sets, i.e., ordering relation play major role in determining the model properties and mechanisms. In this sense it is similar to formal concept analysis (FCA) [Gan99]. One general difference between CoM and FCA is that formal concepts in FCA are defined from the incidence relation which is analogous to primitive semantics in CoM.

Concept structure in CoM makes this approach similar to ontologies [Fen04]. Ontologies is an approach to data and knowledge representation using concepts and relationships between them which allows also reasoning about the elements of the problem domain. In comparison to CoM, ontologies are more complex and involve more basic notions and mechanisms. One feature of CoM is that it provides two orthogonal structures for modelling – physical and logical, and it is based on ordering relation while ontologies hierarchies of classes which are similar to object-oriented approach.

# 7 Conclusions

In this paper we described a new approach to data modelling which incorporates a number of novel aspects. The main of them is the notion of nested ordered set which generalizes conventional ordered sets and is the formal basis for the two-level concept-oriented model. Another novel aspect is the mechanism of syntactic constraints which allows us to restrict the structure of child elements from parent elements of a nested ordered set. Our next contribution is the development of the notion of data semantics which allows us to assign a meaning to the whole model and then treat and manipulate it as one construct. We also define operations with data such as projection and de-projection, developed the mechanism of constraint propagation and studied the notion of dependencies and inference. These formal properties can be applied to very different data modelling problems like grouping and aggregation or multi-dimensional analysis.

The concept-oriented model has many features of the exiting models. It allows the modelling to describe hierarchies like in the hierarchical data model. It supports data connectivity as it is done in the network model. It can use relational operations to manipulate data as it is done in the relational model And it provides facilities similar to multi-dimensional databases. Another amazing feature of CoM is that it is able to resolve the impedance mismatch [Coo06] existing between data manipulation and programming language abstractions because programmers normally tend to encapsulate business logic into object while most of data is stored in a database. CoM is based on the same principles as the concept-oriented programming and hence these approaches are very similar. Half of CoM – physical structure of identity mode – is precisely what is studied in CoM. Thus we can manipulate data elements in CoM as if they were objects in CoP. CoM in this sense simply adds data semantics and semantic operations to normal objects.

Like for any emerging technical trend, we described only main principles and mechanisms. Therefore many of them can probably change in future and many new mechanisms and features will be developed. In particular, we are going to work in the direction of closer integration of CoM with CoP by developing languages for data and code description.